\documentclass[fleqn, usenatbib]{mnras}

\usepackage[T1]{fontenc}
\usepackage{ae,aecompl}

\usepackage{graphicx}	
\usepackage{amsmath}	
\usepackage{amssymb}	
\usepackage{natbib}

\usepackage[usenames, dvipsnames]{color}

\usepackage[caption=false]{subfig}

\usepackage{hyperref}
\usepackage{multirow}

\title[Rapidly quenched galaxies in clusters at $0.5<z<1.0$]{The enhancement of rapidly quenched galaxies in distant clusters at 0.5<\textit{z}<1.0}

\author[M. Socolovsky et al.]
{Miguel Socolovsky,$^{1}$\thanks{E-mail: miguel.socolovsky@nottingham.ac.uk}
Omar Almaini,$^{1}$
Nina A. Hatch,$^{1}$
Vivienne Wild,$^{2}$
\newauthor David T. Maltby,$^{1}$
William G. Hartley,$^{3}$
Chris Simpson$^{4}$
\\
$^{1}$School of Physics and Astronomy, University of Nottingham, Nottingham NG7 2RD, UK\\
$^{2}$School of Physics and Astronomy, University of St Andrews, North Haugh, St Andrews, KY16 9SS, UK\\
$^{3}$Department of Physics and Astronomy, University College London, 3rd Floor, 132 Hampstead Road, London NW1 2PS, UK\\
$^{4}$Gemini Observatory, Northern Operations Center, 670 N.~A`ohoku Place, Hilo, HI 96720-2700, USA
}

\date{Accepted 2018 January 26. Received 2018 January 11; in original form 2017 October 18.}

\pubyear{2018}

\begin{document}
\label{firstpage}
\pagerange{\pageref{firstpage}--\pageref{lastpage}}
\maketitle

\begin{abstract}
We investigate the relationship between environment and galaxy
evolution in the redshift range $0.5<z<1.0$.  Galaxy overdensities are
selected using a Friends-of-Friends algorithm, applied to deep
photometric data in the Ultra-Deep Survey (UDS) field. A study of the
resulting stellar mass functions reveals clear differences between
cluster and field environments, with a strong excess of low-mass
rapidly quenched galaxies in cluster environments compared to the
field. Cluster environments also show a corresponding deficit of
young, low-mass star-forming galaxies, which  show a sharp radial
decline towards cluster centres. By comparing mass functions and radial distributions, 
we conclude that young star-forming galaxies are rapidly quenched as they enter overdense
environments, becoming post-starburst galaxies before joining the red sequence.
Our results also point to the existence of two
environmental quenching pathways operating in galaxy clusters,
operating on different timescales. Fast quenching acts on galaxies
with high specific star-formation rates, operating on
timescales shorter than the cluster dynamical time ($<1$~Gyr). In contrast, slow 
quenching affects galaxies with moderate specific star-formation
rates, regardless of their stellar mass, and acts on longer
timescales ($\gtrsim1$~Gyr). Of the cluster galaxies in the stellar mass range  
$9.0<\log(M_∗/M_\odot)<10.5$ quenched during this epoch, we find that 73\% were 
transformed through fast quenching, while the remaining 27\% followed the slow quenching route.
\end{abstract}




\section{Introduction}

Galaxy properties, such as morphology and star formation activity,
correlate with both environment
\citep{dressler1980,kauffmann2004,balogh2004,linden2010,haines2015}
and the stellar mass of the galaxy
\citep{wel2008,bamford2009}. Massive galaxies and those in dense
environments are predominantly spheroidal and quiescent, whereas lower
mass and field galaxies are mainly disc-dominated and
star-forming. Whilst these trends are most prominent in the present-day Universe, 
it has been shown that the preference for quiescent galaxies to reside in dense 
environments persists until at least redshift $z\sim1.5$ \citep{cooper2007, chuter2011}.

\citet{peng2010} compare the stellar mass and environment of galaxies
with their star formation rate (SFR), and conclude that there are two
separate quenching processes that cause galaxies to cease forming
stars. They call these processes ``environmental quenching'' and ``mass
quenching''. The efficiency of environmental quenching depends on the
environment of a galaxy, such that galaxies in high density environments are more likely to be quenched. Independently, the efficiency of mass quenching 
correlates with the stellar mass of the galaxy, such that more massive
galaxies are more likely to be quenched. Additionally, there is 
morphological quenching \citep{martig2009}, in which the 
structure of the galaxy changes first, leading to a more stable 
configuration which prevents gas from collapsing into stars. However, 
the physical processes that are responsible for these quenching pathways 
remain unclear.

The most popular mechanisms used to explain mass quenching include active galactic nuclei (AGN) feedback \citep{hopkins2005,best2005}, starburst-driven winds \citep{diamond2012} and ``hot halo'' shock heating \citep{dekel2006}.  Interactions between the intracluster or intergroup medium and galaxies, such as ram pressure stripping \citep{gunn1972} and strangulation \citep{larson1980}, are often invoked to explain environmental quenching, as are galaxy-galaxy interactions, such as harassment, mergers and tidal interactions. By measuring the timescale and efficiency of mass and environmental quenching, we can gain insight into where and when these processes act and which is the most important.

Several studies have investigated the timescale of environmental quenching. Semi-analytic models of galaxy formation required gas to be removed on long timescales ($\sim3$--7~Gyrs) to explain the fraction of passive satellites in clusters \citep{font2008, kang2008, weinmann2010, mcgee2011, delucia2012, wheeler2014}. However, the rarity of transitional galaxies can only be explained if the quenching of star formation is rapid \citep{muzzin2012, muzzin2014, wetzel2012, mok2013}. Both observational constraints can be satisfied by a delayed-then-rapid quenching model \citep{wetzel2013}. In this model galaxies experience a delay between the moment they become satellites and when their SFR starts to decline. This time delay can span over $2$--4~Gyrs, but once the SFR begins to decline quenching occurs quickly ($<0.8$~Gyrs). 

One approach to understanding the mechanisms responsible for quenching star formation in galaxies is to examine transitional galaxies. Post-starburst galaxies (PSBs), also referred to as ``k + a'' galaxies, are rare but valuable examples of galaxies caught in transition. Star formation in these galaxies has been rapidly truncated within the past $10^9$~years. They exhibit a red spectral energy distribution (SED), but contain a residual population of A-stars that were born during the starburst phase \citep{dressler1983, wild2009}. These recently quenched galaxies may hold the key to understanding which processes are responsible for environmental and mass quenching.

Until recently, it was very difficult to identify PSBs at $z>0.5$ in large numbers \citep{yan2009, vergani2010, muzzin2014}. The known sample of PSBs was limited because the PSB phase lasts only a short time, and large spectroscopic samples of optically-faint red galaxies are required to identify them. Recently, a new galaxy classification method has  been developed by \citet{wild2014} that only requires photometry. This method, based on a principal component analysis (PCA) of the photometry, has proven effective at classifying SEDs and allows for the identification of large samples of rare galaxies, such as PSBs. This method was verified by \citet{maltby2016} who spectroscopically confirmed that 19 out of 24 ($\sim80$\%) photometrically-selected PSB candidates show genuine PSB features.

In this paper we investigate star-forming, passive and PSB
galaxies in clusters and groups at $0.5<z<1$ to understand the
mechanisms responsible for environmental quenching during this
period. In Section~\ref{sec:data} we describe our data and galaxy
classification method. In Section~\ref{sec:method} we describe our method for 
identifying clusters. We note that our photometric method identifies only
galaxy cluster and group candidates, but we nevertheless refer to them
as ``clusters'' throughout the rest of the paper.  In
Section~\ref{sec:clustersuds} we compare our cluster sample with
previous studies of clusters in the same field. We present our results
in Section~\ref{sec:results} and discuss their significance in
Section~\ref{sec:discussion}. Finally, our conclusions are listed in
Section~\ref{sec:conclusions}. Throughout this paper we use AB
magnitudes and we assume $\Lambda CDM$ cosmology with the following
set of parameters: $\Omega_M=0.3$, $\Omega_\Lambda=0.7$ and
$H_0=100~h~\text{kms}^{-1}\text{Mpc}^{-1}$ with $h=0.7$.

\section{Data sets and galaxy classification}
\label{sec:data}

\subsection{Galaxy catalogue}

We use the $K$-band selected galaxy catalogue described in
\citet{hartley2013}. This catalogue is based on the 8th data release
of the Ultra Deep Survey (UDS; Almaini et al., in preparation) which
covers an area of 0.77 $\text{deg}^2$ to $5\sigma$ depths of $J$=24.9,
$H$=24.2 and $K$=24.6.  The infrared imaging is complemented by deep
optical imaging from the Subaru \textit{XMM--Newton} Deep Survey
(SXDS; \citealp{furusawa2008, ueda2008}),  reaching $5\sigma$ depths
of  $B=27.6$, $V=27.2$, $R=27.0$, $i'=27.0$ and $z'=26.0$. In
addition, our catalogue includes $U$-band imaging 
from the Canada-France-Hawaii Telescope (CFHT) to a $5\sigma$ depth of 
$U=26.75$, and near-infrared
data from the $Spitzer$ Legacy Program (SpUDS)
($\left[3.6\right]=24.2$ and $\left[4.5\right]=24.0$ at
5$\sigma$). The total combined survey area, after masking bright stars
and other features, is $\sim 0.62$ square degrees.

Stars are removed according to the criteria described in
\citet{simpson2013}. The catalogue is limited to $K$<24.3 to ensure
$95\%$ completeness and the resulting catalogue consists of 23,398
galaxies at $0.5<z<1.0$.

\subsection{Photometric redshifts and stellar masses}

Photometric redshifts were derived by \citet{simpson2013} using the
{\sc eazy} photometric-redshift code \citep{brammer2008}, fitting
template spectra to the $U$, $B$, $V$, $R$, $i'$, $z'$, $J$, $H$, $K$,
$3.6\mu \text{m}$ and $4.5\rm\mu m$ photometry.  The
photometric redshifts were tested against $\sim\!1500$ spectroscopic
redshifts from the UDSz (ESO Large Programme, Almaini et al., in prep)
and $\sim3500$ archival redshifts from the literature
\citep{simpson2012}. The resulting normalised median absolute
deviation ($\sigma_\text{NMAD}$) of $z_{\text{phot}}-z_{\text{spec}}$
is $\sigma_\text{NMAD}\sim0.023$.

The stellar masses of the galaxies were computed by
\citet{simpson2013} by fitting a grid of synthetic SEDs to the 
11-band photometry assuming a \citet{chabrier2003} initial mass function
(IMF). The redshift of each galaxy was fixed to the spectroscopic
redshift, if known, otherwise derived properties were based on the photometric redshift.

\subsection{Galaxy Classification and stellar ages}
\label{sec:classification}

We use the galaxy classifications obtained from the PCA analysis
described in \citet{wild2016}, which builds on  the sample
outlined in \citet{wild2014}.  We refer the reader to those papers for
a detailed description of the technique, but we provide a brief
overview below, and  define the various galaxy
subclasses that are used in our work.

The aim of the PCA method was to characterise a broad range of galaxy
spectral energy distributions (SEDs) in a concise manner. It was found that a
linear combination of three base SEDs (``eigenspectra'')  was sufficient to 
describe the range of galaxy SEDs.
The linear coefficients describing the contribution of each eigenspectrum to 
a given galaxy SED is termed a ``supercolour'' (SC).

The supercolour eigenvectors were determined using a grid of 44,000
model SEDs from the stellar population synthesis models of 
\citet{bruzual2003}, using stellar populations with stochastic
star formation histories. These model SEDs are convolved with the
corresponding photometric filters before the PCA is applied. Properties such as $r$-band 
light-weighted stellar ages, sSFRs and metallicities are obtained directly from these models.
It was found that only three eigenvectors are required to characterise
$>$99.9\% of the variance in our model SEDs. Supercolour SC1 alters
the red-blue slope of the SED and traces the $R$-band weighted mean
stellar age or sSFR. Supercolour SC2 modifies the strength of
the Balmer  break region, and traces the fraction of the
stellar mass formed in bursts during the last billion years (burst
fraction), and also correlates with metallicity. Supercolour SC3 also
controls the shape of the SED around $4000$~\!\r{A} and helps to break
the degeneracy between metallicity and burst fraction.

\begin{figure}
 	\begin{center}
 		\includegraphics[width=0.5\textwidth]{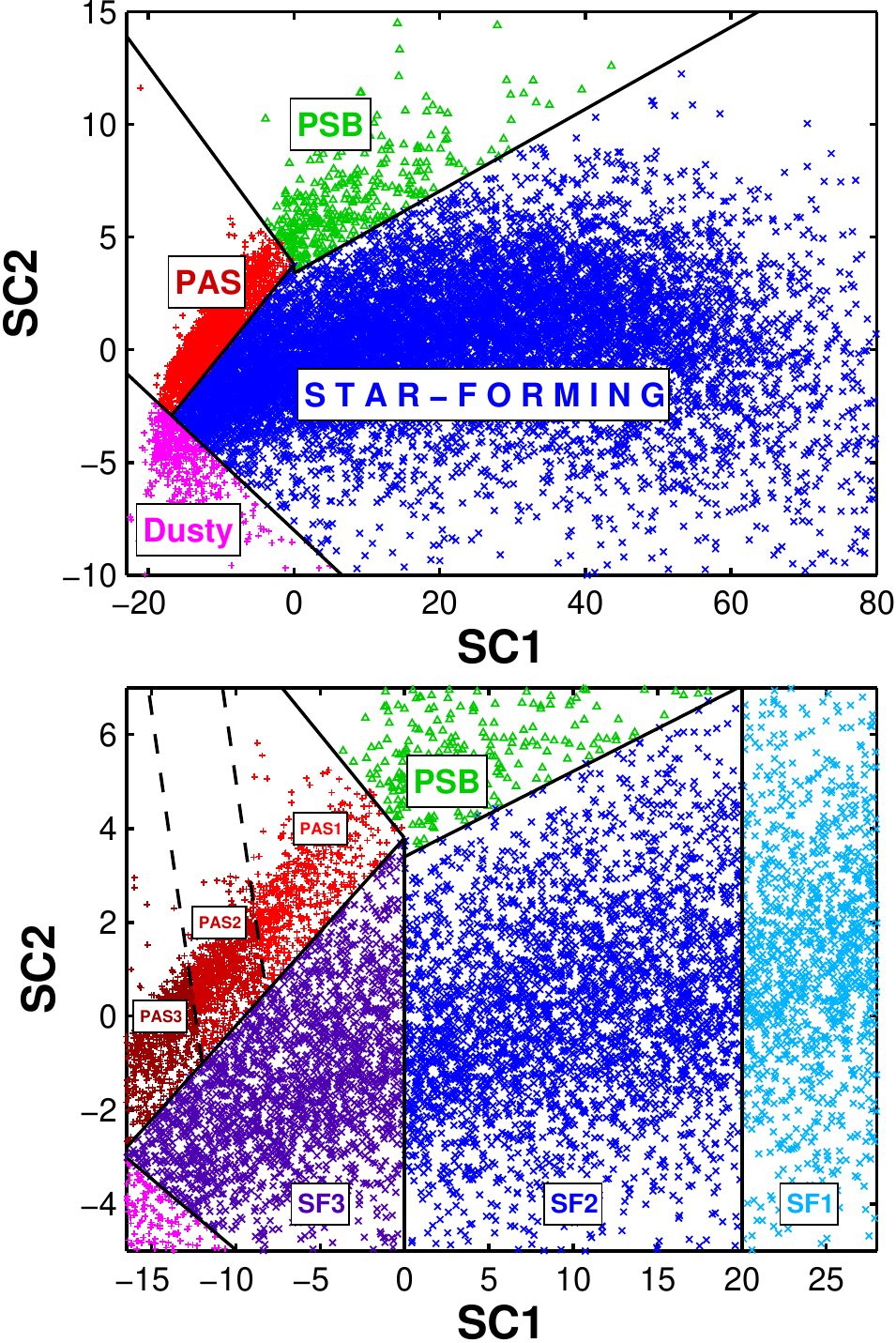}
		\vspace{-10pt}
 		\caption{Top panel: the SC1--SC2 diagram for the
                  galaxies in our sample, based on the PCA
                  classification described in
                  \citet{wild2014}. Galaxies belonging to different
                  populations are represented in different
                  colours. Solid black lines demarcate the borders
                  between the main SC populations. Bottom panel: zoom
                  in of the same diagram showing the sub-populations
                  described in
                  Section~\ref{sec:classification}. Dashed black lines
                  delimit the divisions of the passive galaxy region
                  by mean stellar age.}
 		\label{fig:SCdiagram}
 	\end{center}
 \end{figure}

Galaxies are classified based on their position in the resulting 
SC--SC diagrams (such as shown
in Fig.~\ref{fig:SCdiagram}). The boundaries between the populations
were determined empirically by comparison to both spectroscopy and model SEDs (see \citealp{wild2014} for
more details), and galaxies are divided into the following categories:
star-forming (SF), passive (PAS), post-starburst (PSB), metal-poor and
dusty galaxies (the last two are excluded from our sample). \citet{wild2014} 
subdivide the SF population into 3  groups of decreasing sSFR: SF1, SF2, and SF3.
For our work, we also split the PAS population into three
populations of increasing mean stellar age, from PAS1 to PAS3. This
dividing line was determined by splitting PAS galaxies along the vector
$(\text{SC1},\text{SC2})=(-5,-2)$. The borders ($\text{SC2} =
-\frac{5}{2}\text{SC1}-20$ and $\text{SC2} =
-\frac{5}{2}\text{SC1}-31$) are chosen so that they evenly split the
PAS population into 3 subgroups. The locations of each of the 7
populations on the SC diagram are shown in Fig.~\ref{fig:SCdiagram}.

In total, our galaxy catalogue consists of 11,625 SF1, 3,486 SF2, 2,055
SF3, 575 PAS1, 793 PAS2, 838 PAS3 and 418 PSBs to a magnitude limit of
$K<24$ and in the range $0.5<z<1.0$. We calculate the 90\% mass
completeness limit for each type of galaxy using the method of
\citet{pozzetti2010}. The mass limits at  $z=1.0$ are
$10^{9.0}\text{M}_\odot$ for SF, $10^{9.5}\text{M}_\odot$ for PAS and
$10^{9.3}\text{M}_\odot$ for PSB galaxy populations.  In addition, in
Section~\ref{sec:method} we use a deeper (unclassified) galaxy sample
to $K<24.3$ for the purposes of identifying galaxy overdensities. For
the deeper sample, the 90\% completeness limit as a function of
redshift is described well by the following second-order polynomial:
$\log(M_*)\geq-0.41z^2+1.76z+8.00$.

As an important caveat, we note that we use the term ``PSB'' to refer
to galaxies within the PSB region of the SC diagram. The majority
(60--80\%) of galaxies in this region of the diagram show spectroscopic
``k+a'' properties \citep{maltby2016}, which means they have recently
been rapidly quenched following significant star formation. As noted in
\citet{wild2016}, however, 
this does not necessarily imply that they all 
underwent a ``starburst'' phase before quenching. Very rapid quenching
following a more extended period ($<3$~Gyr) of star formation may also produce
these spectral features.

As a further caveat, we note that spectroscopic confirmation is so far
confined to brighter galaxies ($K<23$), while a large fraction of our
PSBs lie at slightly fainter limits ($23<z<24$). Based on their SEDs,
however, we have no reason to believe that the fainter PSB candidates
show different characteristics, and they populate the PSB region of
the SC diagram in the same way as the brighter
counterparts. Additionally, we note that \citet{maltby2016} exclude galaxies 
with $W_{[\text{OII}]}<-5$~\!\r{A} to rule out PSB candidates with
significant ongoing star formation. We acknowledge that galaxies with
no significant [OII] have been found with residual H$_\alpha$ emission
\citep{yan2006}, but the lack of [OII] together with strong higher
order Balmer absorption lines (i.e. H$_\beta$, H$_\gamma$ and
H$_\delta$) is considered sufficient to rule out significant ongoing 
star formation \citep{goto2003, tran2003, blake2004}.

\section{Cluster Detection Method}
\label{sec:method}

We use a Friends-of-Friends (FoF) algorithm \citep{huchra1982, geller1983, merchan2005} to locate cluster and group candidates in the UDS. For brevity, we refer to candidate groups and clusters as ``clusters'' hereafter. The FoF method is characterised by three parameters: two linking distances, projected ($d_{\text{link}}$) and along the line of sight ($z_{\text{link}}$), and a detection threshold ($N_{\text{min}}$), which is the number of member galaxies per structure. 
The algorithm starts by selecting one galaxy at $[\mathbf{r_0}, z_0]$ from the catalogue which has not been assigned to any structure. All other galaxies fulfilling $\left|\mathbf{r_0}-\mathbf{r_i}\right|\leq d_{\text{link}}$ and $\left|z_0-z_i\right|\leq z_{\text{link}}$ are then designated as ``friends''. The terms $\mathbf{r}$ and $z$ correspond to the position on the sky and redshift, respectively. The method is iterative and continues searching for friends of the friends until no remaining galaxy fulfils the conditions. The structure is classified as a cluster candidate if the number of linked galaxies is greater than $N_{\text{min}}$.

\subsection{Optimising the FoF algorithm}

The completeness and contamination rates of the cluster sample strongly depend on the parameters  $d_{\text{link}}$, $z_{\text{link}}$, and $N_{\text{min}}$. We optimised these parameters to maximise the completeness of the cluster sample whilst also ensuring the cluster sample has no more than 5\% contamination.

To estimate the contamination rate we ran the FoF algorithm on a mock galaxy catalogue using a range of FoF parameters. The mock catalogue had the same number, mean density, and redshift distribution of galaxies as in the UDS, but the RA and Dec were randomised so that the mock catalogue did not contain any groups or clusters. The contamination rate is defined as:
\begin{equation}
q_{\text{cont}} = \frac{N_{\text{mock}}}{N_{\text{UDS}}} 
\end{equation}
where $N_{\text{mock}}$ is the number of clusters detected in the mock catalogue, and $N_{\text{UDS}}$ is the number of clusters detected in the UDS using the same FoF configuration. 

To determine the completeness rate, we injected mock clusters into the UDS catalogue and then attempted to recover them with the FoF algorithm. Mock clusters are constructed as  $N_{\text{sim}}=20$ galaxies randomly distributed within an aperture of radius $R_{\text{sim}}=0.8$~Mpc. Each galaxy is assumed to have a stellar mass of $M_* = 10^{10}\text{M}_\odot$. These simplistic mock clusters result in a conservative estimate of the completeness as real clusters are typically more centrally concentrated, and therefore are easier to detect with a FoF algorithm. All mock clusters are placed at $z_{\text{sim}} = 0.75$, and redshift errors for each galaxy are simulated by randomly sampling a Gaussian distribution of dispersion equal to the photometric redshift uncertainty, $\sigma_\text{z} = (1+z)0.023$. 

We injected $100$ mock clusters in low density regions of the UDS to prevent the mock clusters from overlapping with each other or with existing structures in the UDS. The FoF algorithm is then used to recover the mock clusters. The threshold for recovering a mock cluster is when at least $80\%$ of the injected galaxies are detected and the offset of the centre of mass is less than a $30\%$ of $R_{\text{sim}}$.
The completeness rate ($q_{\text{comp}}$) is defined as the ratio between number of successfully recovered clusters and the number of mock clusters injected into the simulation. A hundred of these simulations are run to obtain the average completeness rate of recovering 10,000 mock galaxy clusters.

We optimise the FoF algorithm by tuning the parameters to maximising the completeness-to-contamination ratio ($r_{\text{comp/cont}}$) while keeping the value of $q_{\text{cont}}$ low. 
\begin{equation}
r_{\text{comp/cont}}=\frac{q_{\text{comp}}}{q_{\text{cont}}}
\end{equation}
The best performing values are: a linking projected distance of $d_{\text{link}}=300$~kpc, and a linking distance along the line of sight of $z_{\text{link}}=40$~Mpc. At a minimum threshold of $N_{\text{min}}=10$ galaxies these parameters yield completeness and contamination rates of $31\%$ and $5\%$, respectively.

\subsection{Limitations of the FoF algorithm}
\label{sec:limitations}

To test the limitations of our FoF cluster finding algorithm we estimated the recovery rate of mock clusters which have a variety of richness ($N_{\text{sim}}$), size ($R_{\text{sim}}$) and redshift ($z_{\text{sim}}$). 
\begin{figure}
	\begin{center}
		\includegraphics[width=0.5\textwidth]{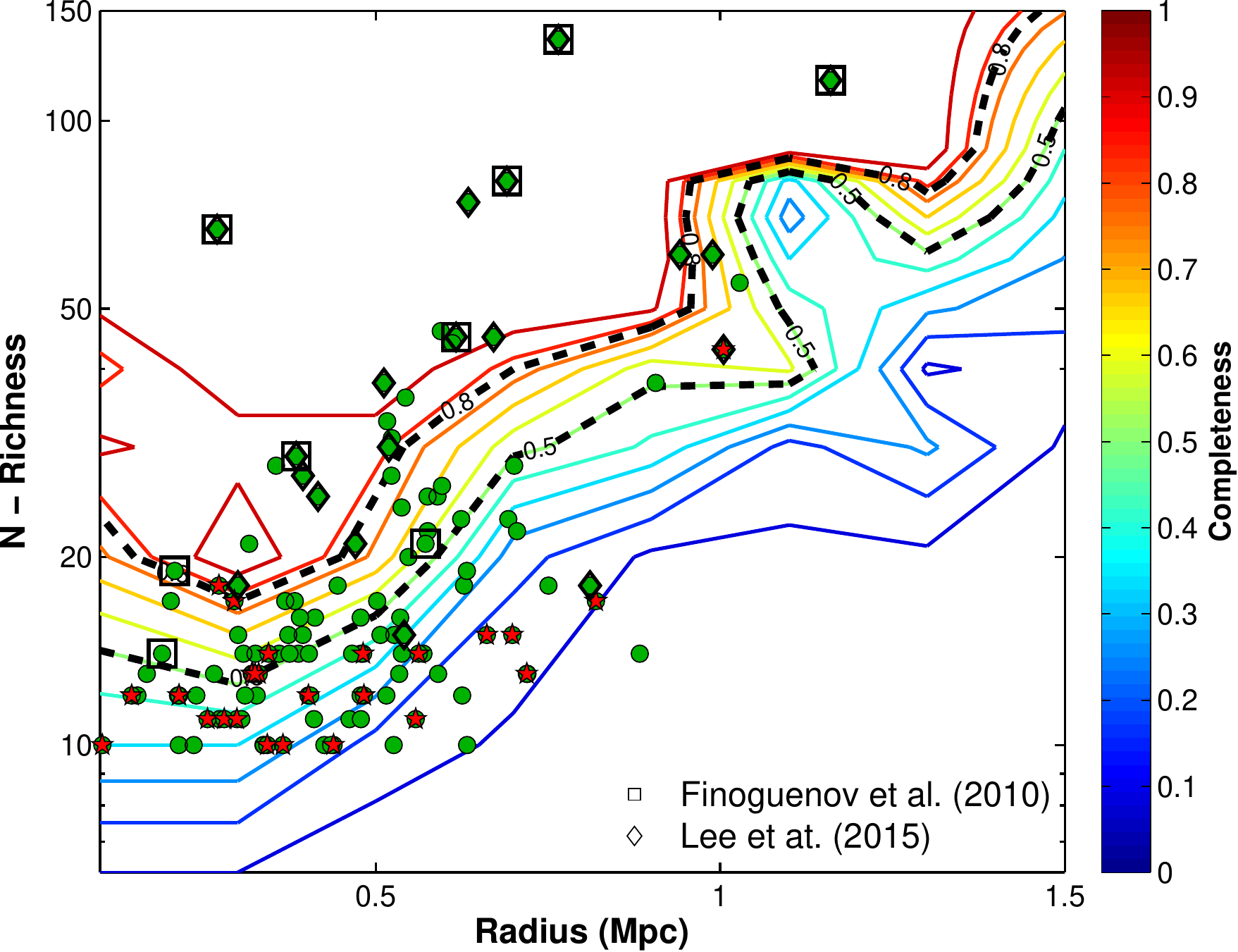}
		\vspace{-10pt}
		\caption{Completeness contours as a function of size and richness of clusters, based on simulated galaxy clusters. Contours of $50\%$ and $80\%$ completeness are highlighted with the thick dashed lines. The dots and stars represent cluster candidates from the UDS. Green dots represent good detections and red stars represent clusters excluded due to a large offset in the centre of mass or low $S/N$ after background subtraction. In addition, cluster candidates coincident with published detections from \citet{finoguenov2010} (boxes) and \citet{lee2015} (diamonds) are included.}
		\label{fig:completeness}
	\end{center}
\end{figure}
Fig.~\ref{fig:completeness} shows that low-richness clusters are only detected if they are also compact. The completeness of our selection method decreases for clusters with small radii, as small deviations in the centre of mass position become significant compared to the size of the cluster. This means that the measured centre of mass for many of the mock clusters deviates from the true centre of mass by more than $30\%$ of $R_{\text{sim}}$. However, this effect becomes important at implausibly small radii ($<100$~kpc), so it does not affect our results.

Fig.~\ref{fig:completeness} shows that our method has low completeness for 
those clusters with fewer than 20 FoF member galaxies. However, this completeness 
is a lower limit because the mock clusters are less likely to be identified by
the FoF algorithm due to the random, rather than centrally
concentrated, spatial distribution of their member galaxies.

\subsection{Cluster centre and effective radius}

We define the projected centre of a cluster as the centre of mass of its FoF members, and its redshift is defined as the median of the photometric redshifts of its FoF members. The effective radius of a cluster, $R_{0.85}$, corresponds to the projected radius that encloses $85\%$ of the stellar mass of the system.

The centre of a cluster can also be defined as the mean or median of the RA and Dec of all FoF members. The  cluster centre should not depend strongly on the definition used, unless the cluster has no well-defined centre. Therefore, we remove 10 clusters from our sample whose measured centroid deviates by more than $30\%$ of $R_{0.85}$ depending on which definition is used (see Fig~\ref{fig:signaltonoise}).

\subsection{Cluster galaxy membership}

The FoF algorithm is optimised to identify clusters in the UDS, but the galaxy membership of these clusters will be incomplete due to photometric redshift errors. To correct for missing galaxies, we define candidate cluster members as all galaxies within a cylinder around the centre of mass of each cluster. Each cylinder has a radius  of $R_{\text{cyl}} = 1$~Mpc, which is the typical size of a galaxy cluster,  and a depth of $\delta z_{\text{cyl}} = 2.5\sigma_\text{z}$, which corresponds to $\sim250$~Mpc in our redshift range. 

The large photometric redshift uncertainties means we must use long cylinders to avoid missing cluster galaxies, but this implies that the cylinders may include a significant fraction of field galaxies, which are considered contaminants. These contaminants can be removed by statistically subtracting the field galaxies expected in each cylinder.

\subsection{Construction of a field galaxy sample}

\begin{figure}
	\begin{center}
		\includegraphics[bb = 1.5cm 6.5cm 20cm 21cm, clip=true, width=0.5\textwidth]{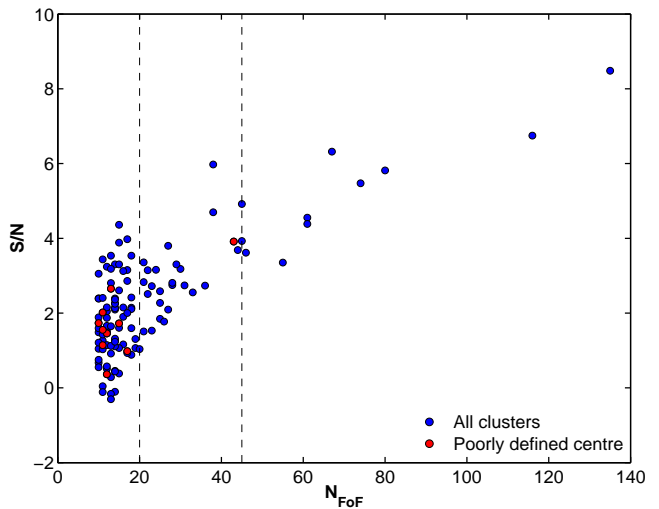}
		\vspace{-10pt}
		\caption{Signal-to-noise (S/N) ratio of the cluster detections as a function of richness of our cluster sample, using the method described on Section \ref{sec:sn}. Dashed lines divide the richness into the three bins we utilise in the following sections. Clusters with poorly defined centres are shown as red dots, which seem to be concentrated in the lowest richness bin ($N<20$ galaxies), making this the most contaminated and unreliable regime.}
		\label{fig:signaltonoise}
	\end{center}
\end{figure}

We construct a sample of field galaxies to remove the field contribution within the cylindrical volume containing the cluster members, and to use as second environment to compare with our cluster sample. 

The field sample is constructed from the UDS. For each cluster cylinder, a field sample is defined as all galaxies in the UDS (which are not candidate cluster members) that lie within the same redshift interval as the cylinder. 
The number of galaxies in the field is then scaled by the ratio of unmasked pixels in the cluster region to the field region, so that the field corresponds to the same volume as the cluster, i.e. a cylinder with radius 1~Mpc and depth 250~Mpc. The rescaled field number count ($N_{Field}^*$) can be expressed as the original number scaled by a normalisation factor, $f$:
\begin{eqnarray}
 N_{Field}^{*} = fN_{\text{Field}} = \frac{n_{\text{cyl}}}{n_{\text{Field}}}N_{\text{Field}}
\end{eqnarray}
where $n_{\text{cyl}}$ is the number of good pixels inside the aperture corresponding to the cylinder and $n_{\text{Field}}$ is the total number of good pixels across the field sample.
Finally, all the separate field regions corresponding to each detected cluster are combined together to produce the total field galaxy sample. We define a field sample for each cluster, but several clusters have similar redshifts so the total combined field sample contains some duplication of UDS galaxies. This duplication amounts to less than 10\% of the total field sample.

\subsection{Signal-to-noise ratio of the cluster detections}
\label{sec:sn}

To determine a quality control for our cluster detections, we define the signal-to-noise ratio of each cluster detection as 
\begin{eqnarray}
 S/N = \frac{N_{\text{cluster}}-fN_{\text{field}}}{\sqrt{\sigma_{\text{cluster}}^2 + \sigma_{\text{field}}^2}} = \frac{N_{\text{cluster}}-fN_{\text{field}}}{\sqrt{N_{\text{cluster}} + f^2N_{\text{field}}}},
\end{eqnarray}
where $N_{\text{cluster}}$ is the number of galaxies in the cylindrical volume around the cluster, $N_{\text{field}}$ is the number of galaxies in the field corresponding to the same redshift interval, and $f$ is the scale factor that resizes the field to the cylindrical volume of the cluster.

Fig.~\ref{fig:signaltonoise} displays the richness (defined as the number of FoF members) and the $S/N$ of our cluster sample. Richer clusters have a higher $S/N$. Only 3\% of clusters with more than 20 members have poorly defined centres, whilst 17\% of clusters with less than 20 member galaxies have poorly defined centres, and 25\% have a $S/N$ lower than unity. Based on both the low $S/N$ and the low completeness rate found in Section~\ref{sec:limitations}, we decide to exclude those clusters with fewer than 20 member galaxies. This ensures a high quality cluster sample, although it significantly reduces the sample size.

\section{Clusters in the UDS}
\label{sec:clustersuds}

The FoF algorithm identifies 37 galaxy cluster candidates at $0.5<z<1.0$ in the UDS field. Eleven cluster candidates contain more than 45 FoF members, whilst 26 have between 20 and 45 members. This results in a sample of 2210 cluster galaxies\footnote{Cluster galaxies are defined as all the galaxies within the cylinder encompassing the cluster.} (of which 98 are classified as PSBs) and 13,837 field galaxies (220 of which are PSBs). We also identify 87 cluster candidates with less than 20 and more than 10 FoF members, but we do not analyse these further as this sample has a high level of contamination.

\begin{figure}
	\begin{center}
		\includegraphics[bb = 1cm 6.5cm 21cm 21.5cm, clip=true, width=0.5\textwidth]{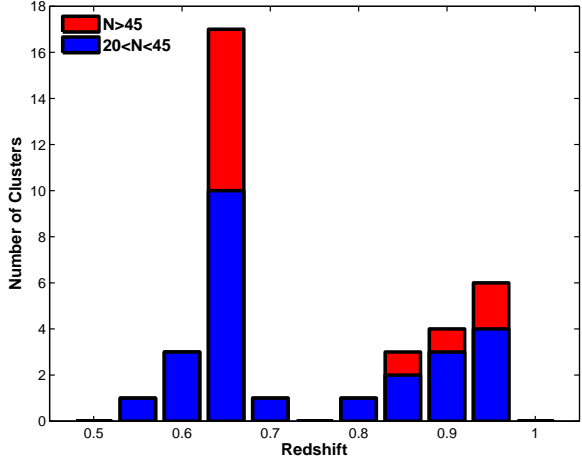}
		\vspace{-10pt}
		\caption{Distribution of detected clusters as a function of redshift. In the histogram red colour indicates clusters with more than 45 members and blue indicates clusters with more than 20 but less than 45 galaxy members.}
		\label{fig:clusters}
	\end{center}
\end{figure}

The catalogue of our cluster candidates is provided in Table~\ref{tab:listofclusters} and their redshift distribution is shown in Fig.~\ref{fig:clusters}. A spike in the redshift distribution of clusters is visible at $z\sim0.65$ due to the presence of a well-known galaxy overdensity, including a massive cluster in the CANDELS-UDS region \citep{geach2007}. These structures are not fragments of the same massive cluster as they appear evenly spread across the UDS field.  Instead, most of these structures are likely to be smaller clusters surrounding the massive cluster, since clusters of galaxies are highly clustered.

\begin{table*}
\centering
\caption{Catalogue of galaxy cluster candidates detected in the UDS using the FoF algorithm. Identification number is provided in column 1, RA and Dec $(2-3)$, photometric redshift $(4)$. Column 5 corresponds to the median spectroscopic redshift of the spectroscopically confirmed clusters (see Section~\ref{spec-confirmation}) and the number of spectroscopic redshifts associated with the structure $(6)$. Three measurements of the richness of the clusters: number of FoF members $(7)$, field subtracted number of galaxies within 1~Mpc from the cluster centre $(8)$ and field subtracted stellar mass within 1~Mpc from the centre $(9)$. Column 10 provides references if the structure has been previously detected. The bottom two rows correspond to clusters that are spectroscopically confirmed despite having fewer than 20 FoF members.}
\label{tab:listofclusters}
\begin{tabular}{cccccccccc}
\hline
ID          & RA       & Dec      & phot-z & median & N($z_{\text{spec}}$) & N$_{\text{FoF}}$ & N$_{\text{Sub}}$ & M$_{\text{Sub}}$ & Reference \\
            & (deg)    & (deg)    &        & spec-z &               &           &  (1~Mpc)  &$\log(M_*)$&        \\
(1)         & (2)      & (3)      &   (4)  & (5)    & (6)           &  (7)      &   (8)     &  (9)      &  (10)  \\
\hline
UDSC01FOF & 34.70321 & $-$5.14147 & 0.546 &       &          & 21   & 23  & 11.9312 & ${}^{b}$        \\ 
UDSC02FOF & 34.28647 & $-$5.07732 & 0.609 &       &          & 22   & 22  & 11.2892 &                 \\
UDSC03FOF & 34.24918 & $-$5.18202 & 0.618 &       &          & 21   & 13  & 11.8515 & ${}^{a}$        \\ 
UDSC04FOF & 34.64570 & $-$4.96700 & 0.620 & 0.589 & 14       & 38   & 46  & 12.0126 & ${}^{a,\,b}$    \\ 
UDSC05FOF & 34.59033 & $-$5.29313 & 0.627 &       &          & 28   & 25  & 11.9037 &                 \\ 
UDSC06FOF & 34.35261 & $-$5.41159 & 0.628 &       &          & 45   & 52  & 12.1130 & ${}^{b,\,c}$    \\ 
UDSC07FOF & 34.42521 & $-$5.46676 & 0.631 &       &          & 25   & 24  & 11.8822 &                 \\ 
UDSC08FOF & 34.18869 & $-$5.14456 & 0.631 &       &          & 45   & 39  & 11.8729 & ${}^{b}$        \\ 
UDSC09FOF & 34.29001 & $-$5.13710 & 0.632 &       &          & 27   & 18  & 11.8189 &                 \\ 
UDSC10FOF & 34.53183 & $-$5.36065 & 0.635 &       &          & 27   & 38  & 11.8080 & ${}^{a,\,b}$    \\ 
UDSC11FOF & 34.67991 & $-$5.38076 & 0.637 &       &          & 28   & 26  & 11.3117 &                 \\ 
UDSC12FOF & 34.28599 & $-$5.42808 & 0.638 &       &          & 55   & 32  & 11.8175 &                 \\ 
UDSC13FOF & 34.58946 & $-$5.38840 & 0.638 &       &          & 38   & 67  & 12.3032 &                 \\ 
UDSC14FOF & 34.39740 & $-$5.22350 & 0.638 & 0.647 & 20       & 135  & 111 & 12.4485 & ${}^{a,\,b,\,c,\,d}$\\ 
UDSC15FOF & 34.54191 & $-$5.25419 & 0.641 & 0.647 & 10       & 74   & 57  & 12.2359 & ${}^{b}$        \\ 
UDSC16FOF & 34.60487 & $-$5.41888 & 0.646 & 0.647 & 13       & 67   & 73  & 12.3414 & ${}^{b,\,c,\,d}$    \\ 
UDSC17FOF & 34.64400 & $-$5.01744 & 0.648 &       &          & 44   & 36  & 11.8114 &                 \\ 
UDSC18FOF & 34.62682 & $-$5.34075 & 0.651 &       &          & 31   & 25  & 11.6202 &                 \\ 
UDSC19FOF & 34.34840 & $-$5.18454 & 0.651 & 0.649 & 10       & 24   & 30  & 11.8912 &                 \\ 
UDSC20FOF & 34.53353 & $-$5.51288 & 0.671 &       &          & 43   & 36  & 11.8701 & ${}^{b}$        \\ 
UDSC21FOF & 34.49045 & $-$5.45092 & 0.674 & 0.695 & 7        & 116  & 79  & 12.3302 & ${}^{b,\,c}$    \\ 
UDSC22FOF & 34.37161 & $-$4.69193 & 0.681 &       &          & 25   & 15  & 11.4853 & ${}^{b}$        \\ 
UDSC23FOF & 34.21696 & $-$5.20876 & 0.814 &       &          & 23   & 21  & 11.8909 & ${}^{a}$        \\ 
UDSC24FOF & 34.52203 & $-$4.73357 & 0.850 &       &          & 30   & 27  & 11.9332 & ${}^{a,\,b}$    \\ 
UDSC25FOF & 34.82970 & $-$5.08690 & 0.872 & 0.872 & 9        & 29   & 30  & 12.1240 & ${}^{b,\,c}$    \\ 
UDSC26FOF & 34.63429 & $-$5.01229 & 0.874 & 0.874 & 31       & 80   & 67  & 12.3855 & ${}^{a,\,b,\,c}$\\ 
UDSC27FOF & 34.36706 & $-$4.70291 & 0.876 &       &          & 26   & 15  & 11.4445 &                 \\ 
UDSC28FOF & 34.71698 & $-$5.35764 & 0.899 &       &          & 46   & 37  & 12.1644 &                 \\ 
UDSC29FOF & 34.27406 & $-$5.16789 & 0.910 &       &          & 20   & 9   & 11.6155 &                 \\ 
UDSC30FOF & 34.76268 & $-$4.70390 & 0.910 &       &          & 36   & 24  & 12.0208 & ${}^{a}$        \\ 
UDSC31FOF & 34.52417 & $-$5.37735 & 0.918 &       &          & 25   & 22  & 11.7250 &                 \\ 
UDSC32FOF & 34.87913 & $-$5.22070 & 0.926 &       &          & 23   & 12  & 11.9276 &                 \\ 
UDSC33FOF & 34.80408 & $-$4.91053 & 0.926 &       &          & 21   & 33  & 11.9349 & ${}^{c}$        \\ 
UDSC34FOF & 34.34259 & $-$5.20107 & 0.937 & 0.918 & 6        & 61   & 49  & 12.0711 & ${}^{a,\,b}$    \\ 
UDSC35FOF & 34.28586 & $-$4.96203 & 0.953 &       &          & 33   & 27  & 11.8483 &                 \\ 
UDSC36FOF & 34.04102 & $-$4.86472 & 0.953 &       &          & 61   & 50  & 12.1284 & ${}^{b}$        \\ 
UDSC37FOF & 34.28933 & $-$4.76095 & 0.957 &       &          & 22   & 33  & 12.0459 & ${}^{a}$        \\
\hline                                                                     
UDSC38FOF& 34.50443 & $-$4.79895 & 0.568 & 0.583 & 14 & 13   & 22  & 11.9007 &          \\ 
UDSC39FOF & 34.39913 & $-$5.07272 & 0.800 & 0.801 & 10 & 13   & 25  & 11.9074 &            \\
\hline                                                                 
\end{tabular}\\ 
${}^a$ Detected by \citet{breukelen2006}, ${}^b$ detected by \citet{lee2015}, ${}^c$ detected by \citet{finoguenov2010}, ${}^d$ detected by \citet{geach2007}
                                                                       
\end{table*}

\subsection{Spectroscopic confirmation of cluster candidates}
\label{spec-confirmation}

To spectroscopically confirm our cluster sample, we utilise more than 6800 spectroscopic redshifts from the UDS field, including 1511 secure redshifts from the UDSz (ESO Large Programme, Almaini et al., in prep) and over 3000 archival redshifts from Subaru FOCAS and AAT 2dF (Akiyama et al. 2010; in prep), VLT VIMOS (Simpson et al. 2010; in prep), AAOMEGA \citep{smail2008} and VIPERS \citep{scodeggio2016}. We classify a cluster as spectroscopically confirmed if it contains at least five spectroscopic galaxies within a cylinder of $\pm1000~\text{kms}^{-1}$ length and 1~Mpc radius \citep{eisenhardt2008}. In addition, the median of the spectroscopic cluster galaxies must not be offset by more than $1\sigma$ from the photometric redshift of the candidate cluster. Eleven of our cluster candidates fulfil these conditions (see Table~\ref{tab:listofclusters}), of which three have not been previously presented in the literature.

\subsection{Comparison with previous studies of clusters in the UDS}

Clusters in the UDS have been located by \citet{finoguenov2010} through the detection of extended \textit{XMM--Newton} X-ray emission; by \citet{breukelen2006} and \citet{lee2015}, who searched for galaxy overdensities in the optical and near infrared photometric surveys, and by \citet{geach2007}, who used low-power radio galaxies as beacons for overdensities. We compare cluster samples derived from these methods with our FoF cluster sample to check the robustness of our detection method. Throughout this comparison, we use our whole sample of cluster candidates with a richnesses greater than 10 FoF galaxies. Although many of the cluster candidates with less than 20 FoF members are likely to be contaminants, some of them are expected to be real clusters, as shown by Fig.~\ref{fig:completeness}.

The two spectroscopically confirmed clusters at $z=0.65$ from \citet{geach2007} are two of the most massive structures we select.
We locate $83.3\%$ $(10/12)$ of the cluster candidates detected by \citet{breukelen2006}, who used an algorithm based on FoF and Voronoi tessellation\footnote{We define a cluster match if the RA and Dec of the cluster centre matches to within 2 arcmin ($\sim1$~Mpc) and $\Delta z\lesssim\sigma_\text{z}$, where $\sigma_\text{z}$ represent the total photometric redshift uncertainty i.e.\ the combination of the literature and our photometric redshift uncertainties. Furthermore, we ignore known or candidate clusters from the literature that fall within masked regions of our catalogue or lie outside our $0.5<z<1.0$ redshift interval.}. However, there seems to be a systematic bias in their cluster redshifts with respect to ours as theirs tend to be systematically lower at $z>0.7$. This offset is probably due to the relatively unreliable photometric redshifts from the UDS DR1 catalogue used by \citet{breukelen2006}, which was much shallower than the DR8 catalogue.
We recover $85.2\%$ $(17/20)$ of the cluster candidates listed in \citet{lee2015}, where they locate clusters as galaxy overdensities in spatial and photometric redshift space.
We also locate $78.5\%$ $(11/14)$ of the X-ray selected cluster candidates in \citet{finoguenov2010}.
The 3 structures that we miss are close to our lower redshift limit at $z=0.514$, $0.517$ and $0.548$.

Two X-ray selected cluster candidates at $z=0.548$ and $z=0.514$ (named SXDF66XGG and SXDF42XGG, respectively in \citealp{finoguenov2010}) may be misclassified groups of X-ray
AGN that are close in projection on the sky. No galaxy excess is detected near either of these cluster candidates. However, three \textit{Chandra} X-ray point sources are located at angular separations of $6.96''$, $8.14''$ and $15.20''$ from the centre of the SXDF66XGG cluster, each of them with a galaxy counterpart within 1 arcsec. Similarly, two X-ray point sources from the Subaru \textit{XMM--Newton} Deep Survey \citep{akiyama2015} are found within $7.94''$ and $10.70''$ from the centre of SXDF42XGG. These two sources have galaxy counterparts offset $1.51''$ and $3.81''$, respectively, from the X-ray source, which is within the \textit{XMM} point-source error circle.

The cluster candidate SXDF24XGG, at $z=0.517$, shows a slight excess of galaxies in our catalogue. We detect the candidate as a group of 5 FoF galaxies when we optimise the algorithm to detect clusters at $z\sim0.5$. When the algorithm is optimised to locate clusters across the redshift range $0.5<z<1.0$ it begins to break down at both redshift extremes, but especially at low redshift. Hence, it is likely that this small cluster is missed by our original detection algorithm.

We conclude that we do not detect all the X-ray cluster candidates from \citet{finoguenov2010} because the presence of one or more X-ray point-sources (AGN) means that some cluster candidates are falsely identified as extended sources due to the low resolution of the \textit{XMM--Newton} data. Furthermore, the X-ray cluster detection method is highly efficient at low redshift where our ability to detect clusters through the FoF algorithm decreases. This is supported by the test simulations shown in Fig.~\ref{fig:completeness} where some X-ray cluster candidates lie in the low completeness regime of our method.

\section{Results}
\label{sec:results}
In this section we compare the properties of galaxies identified in
our 37 candidate galaxy clusters with those identified in the field,
focussing on the redshift range $0.5<z<1.0$. The ``cluster'' sample
consists of galaxies identified in overdense regions containing at
least 20 members, linked by the Friends-of-Friends algorithm, as
described in Section~\ref{sec:method}.

In Section~\ref{sec:quenchinginclusters} we compare the PCA
supercolours for the cluster and field samples, while in
Section~\ref{sec:SMF} we compare the stellar mass functions. In
Section~\ref{sec:radialplots} we investigate the radial distribution of galaxies for the cluster populations.

\subsection{Cluster and field galaxy populations}
\label{sec:quenchinginclusters}

\begin{figure*}
	\begin{center}
		\includegraphics[width=0.9\textwidth]{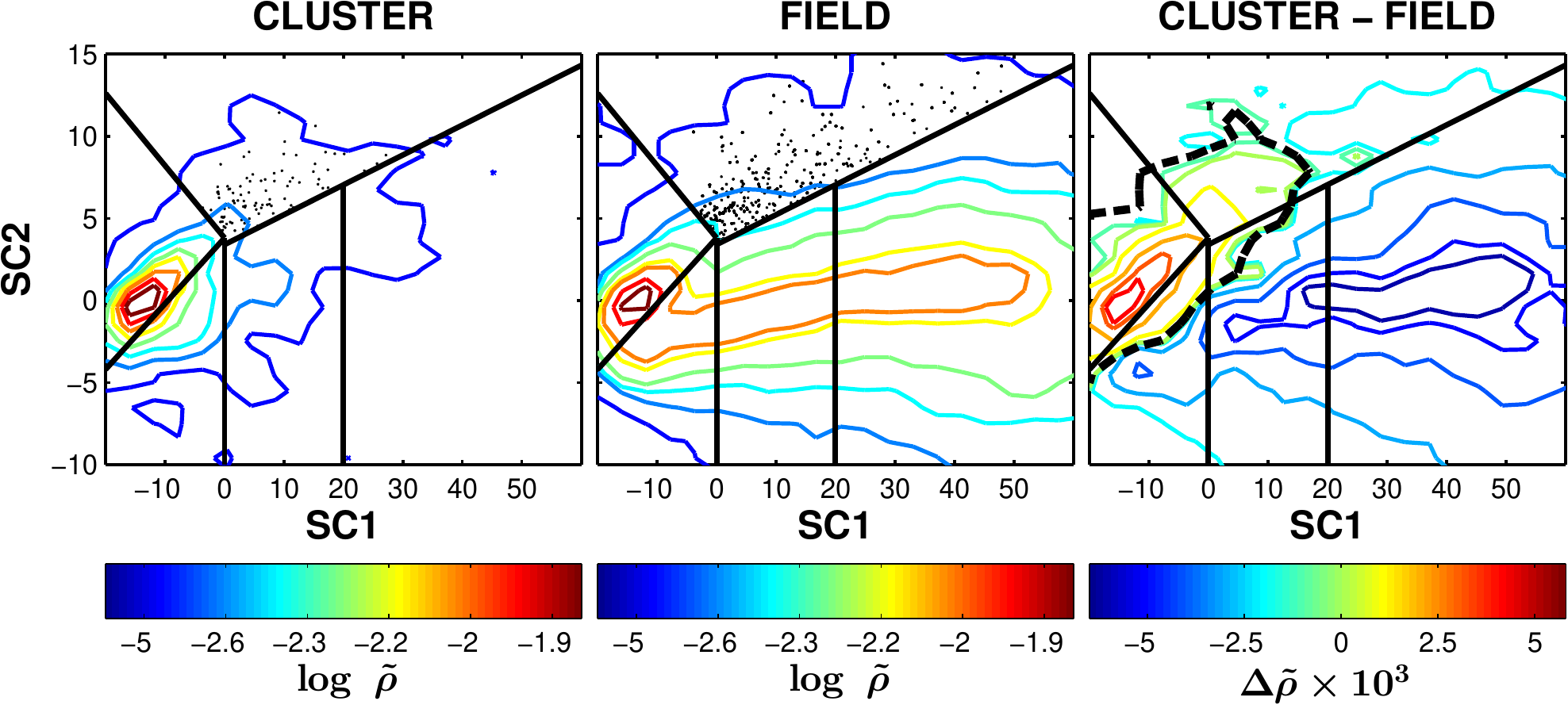}
		\vspace{-5pt}
		 \caption{The distribution of UDS galaxies at $0.5<z<1.0$ across the SC-space. Straight solid black lines represent the boundaries between the different galaxy populations and black dots the PSBs in the sample. Colour contours show the number of galaxies per bin normalised by the total number of galaxies in the diagram, where the bin size is $\Delta \text{SC1}\times\Delta \text{SC2} = 4\times 1$. The panel on the left shows the distribution of cluster galaxies (note it has been field subtracted). The central panel shows the distribution of field galaxies. The right-hand panel shows the difference between cluster and field densities, with the dashed black contour representing the regime where field and cluster have the same density.}
		\label{fig:SCdistribution}
	\end{center}
\end{figure*}

In Fig.~\ref{fig:SCdistribution} we present the number density of
galaxies across the SC1--SC2 diagram for our candidate galaxy clusters
and  the field. For the cluster sample, the densities across the SC diagram are
obtained after subtracting the corresponding values for the field
(correcting for the volumes sampled), to correct for the
contamination from field galaxies in the cluster
volumes. We find significant differences between the cluster and field
populations, which are emphasized in the final panel, which displays
the difference between the cluster and field regions.

We observe that galaxies in clusters are, in general, more evolved
than those in the field. The differences are reflected in the overall
shift of cluster galaxies towards the left side of the SC-diagram,
producing an enhancement of the quiescent galaxies (PAS) and star-forming galaxies in the SF3
class, characterised by their high mean stellar ages and low sSFRs.
Following the same trend, there is a lack of young star-forming
objects in clusters (at high values of SC1). The SF1 class, with the 
highest sSFR,  is common in the field but rare in clusters.

There are PSBs in both environments, but their distributions over the
SC-space is significantly different. While PSBs in the field are found
to be widespread over the upper region of the diagram, their
counterparts in dense environments only populate the area closest to
the border with the PAS population ($\text{SC2}<10$). A two-sample
Kolmogorov-Smirnov (KS) test applied only to SC2, rejects the null 
hypothesis that the field and cluster PSBs are drawn from the same 
underlying distribution (giving a probability of $1.45\times10^{-6}$). 
This difference may suggest that PSBs are formed via different mechanisms, 
depending on their environment. We explore this result and its possible
implications in Section~\ref{sec:PSBinclusters}.

\subsection{Mass Functions of clustered galaxies vs. the field}
\label{sec:SMF}

Stellar mass functions can provide further information on the
evolution of galaxies and, in particular, about the range of masses
affected by environmental quenching. In this section we present the
stellar mass functions of SF, PAS and PSB galaxies split by
environment. Additionally, we split the SF category by decreasing sSFR
(SF1, SF2 and SF3) and the PAS sample by increasing mean stellar age
(PAS1, PAS2, PAS3), using the classification boundaries defined in 
Section~\ref{sec:classification}.

\begin{figure*}
	\begin{center}
		\includegraphics[width=.9\textwidth]{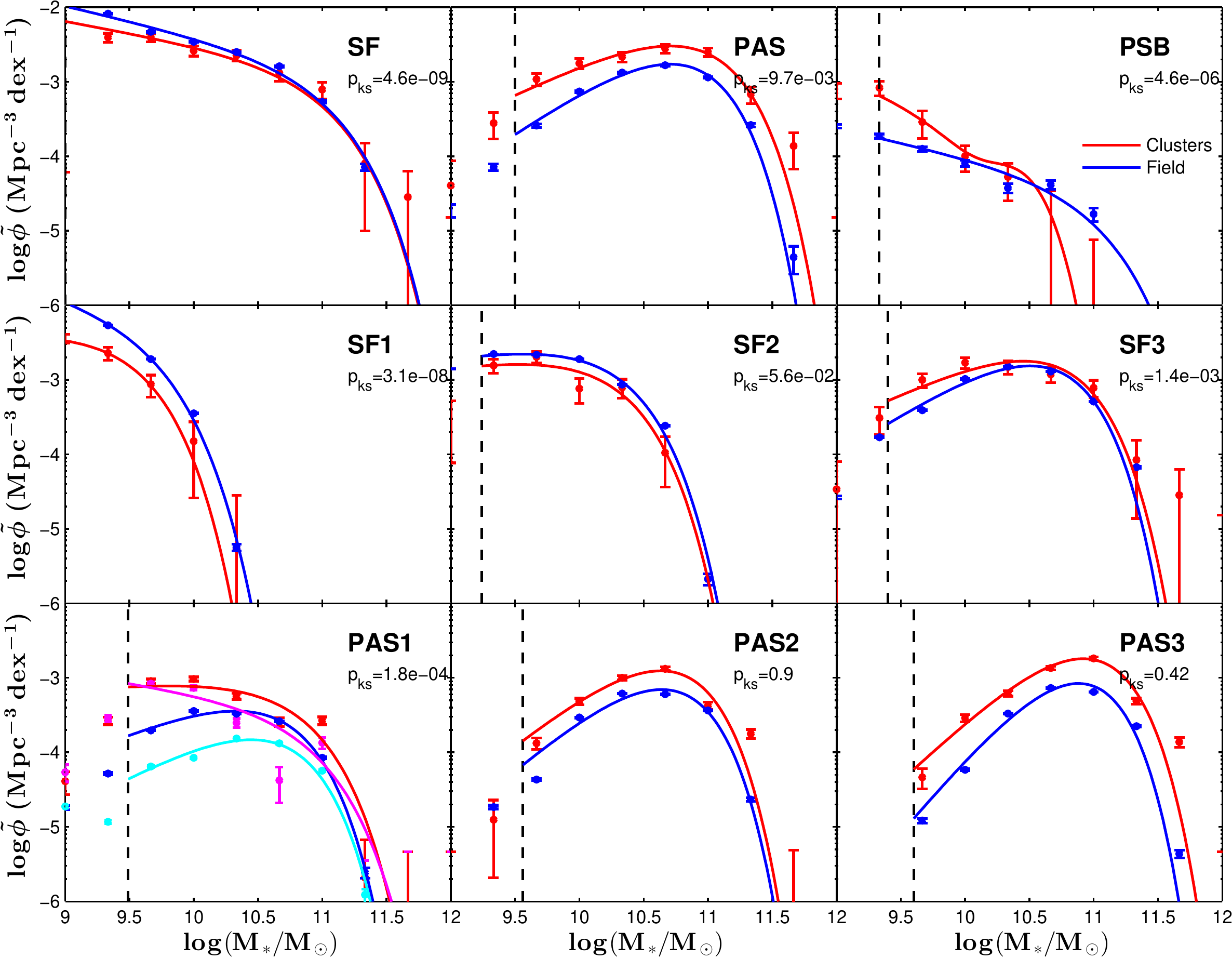}
 		\vspace{-5pt}
		\caption{Stellar mass functions of galaxies in
                  clusters (red) and the field (blue) at
                  $0.5<z<1.0$. The cluster mass functions are
                  normalised so that the total (integrated) density of
                  galaxies matches the field. The first row
                  corresponds to the three main galaxy populations:
                  SF, PAS and PSB, from left to right. The second and
                  third rows represent the mass functions of the three
                  sub-populations of the SF and PAS categories,
                  respectively, ordered from young to old (from left
                  to right). In the panel corresponding to the PAS1
                  population, the stellar mass functions of galaxies
                  quenched during the redshift interval $0.5<z<1.0$
                  are represented with magenta and cyan lines
                  for cluster and field, respectively. The vertical
                  dashed black line indicates the $90\%$ mass
                  completeness limit. Additionally, each panel shows
                  the probability that the field and cluster samples
                  are drawn from the same underlying population,
                  according to a KS test, as applied to the sample
                  before statistical background subtraction.}
		\label{fig:massfunction1}
	\end{center}
\end{figure*}

The stellar mass functions shown in Fig.~\ref{fig:massfunction1} are
computed using the cluster and field samples. Since the cluster total
densities are arbitrary, given that the volume of the cylinder is
chosen artificially, the cluster mass functions are normalised so that
the total density (of all galaxies) matches the total density in the
field. This allows us to compare the shapes of the mass functions
across environments and populations, but  implies that a comparison
of  normalisations (i.e.\ total densities) is only meaningful
within the same environment. Although the normalisation is arbitrary,
all densities are offset by the same amount from the true
 cluster density; we parametrise this offset by
introducing the quantity $\xi$ whose exact value is unknown to us.
\begin{equation}
\xi = \frac{total~density~in~clusters}{total~density~of~the~field}
\end{equation}

Cluster galaxy mass functions are computed using the cluster sample
described in  Section~\ref{sec:clustersuds},
consisting of 37 candidate clusters at $0.5<z<1.0$ with more than 20
members linked by the FoF algorithm. The field mass function
is subtracted in order to remove background contamination. We fit
simple Schechter functions to all our mass functions except to the
cluster PSBs, to which we fit a double Schechter mass function, with two
power laws and one exponential \citep{pozzetti2010}. This is because
we believe the cluster PSB class comprises two different populations;
one which is identical to that observed in the field and one that is
produced by environmental quenching \citep[see also][]{wild2016}. The
list of fitted Schechter parameters is given in
Table~\ref{tab:SchechterParam}. Fits were performed using a Maximum
Likelihood method using unbinned data \citep{marshall1983}.

The stellar mass functions of the three main populations show
significant differences as a function of environment, with PSBs
showing the largest difference between clusters and the field. The
probability (p-value) of both populations being drawn from the same
distribution according to a KS test is
$p_{\text{KS}}=4.2\times10^{-6}$. The stellar mass function of this 
population suggests that they are very strongly clustered, as the 
number density is more than $3~\xi$ times larger in clusters
than in the field. The shape of the mass function is also very
different; PSBs in clusters are predominantly low-mass galaxies
($M<10^{10.5}~\text{M}_\odot$) while in the field the range of masses
is broader.

The PAS population also shows a strong environmental
dependence. Passive galaxies are more abundant in clusters, as
expected, with $2.5~\xi$ times the density of the field. More
interesting is the different shape of the passive galaxy mass function
in clusters with respect to the field, with evidence for an excess of
low-mass galaxies; we reject the null hypothesis that the populations
are drawn from the same underlying distribution at 
significance $p_{\text{KS}}=9.9\times10^{-3}$. Furthermore, we see
that this excess is mainly produced by the ``younger'' passive  galaxies (i.e. the most
recently quenched), with PAS1  presenting
$p_{\text{KS}}=1.4\times10^{-4}$ between field and cluster.

The  SF population also  presents a stellar
mass distribution that depends on environment
($p_{\text{KS}}=3.6\times10^{-9}$), with a deficit of 
low-mass galaxies in cluster environments. Unlike the PAS and PSB populations, the
overall density in the field is $\sim1.2~\xi$ times higher than in clusters,
which indicates that SF galaxies have no preference for dense
environments. Some studies have found the opposite trend, suggesting
a high fraction of star-forming galaxies in dense environments at $z\sim 1$
\citep{elbaz2007, cooper2008}. However, these were conducted using
optical galaxy selection, which has been shown to be strongly biased
towards blue star-forming galaxies at high redshift. With the rise of
near-infrared surveys, it was found that the star
formation--density relation was in place already at $z\sim1$--$1.5$ \citep{williams2009, chuter2011}. 

Studying the three SF sub-populations we find a strong dependence of
quenching with both sSFR and stellar mass. The population with the
highest sSFR (SF1) is found to be strongly suppressed in
clusters. This suppression is also mass-dependent and is more
efficient at low stellar masses; a KS test rejects the null hypothesis
that mass distributions in clusters and the field are drawn from the
same underlying population ($p_{\text{KS}}=3.1\times10^{-8}$).
For the intermediate class (SF2) we find a slight suppression in the relative number density
in cluster environments, but no evidence for a change in the shape of the mass function.
In
contrast to SF1 galaxies, the relative abundance of the SF3 population appears to be
enhanced in cluster environments, with evidence for an excess of
low-mass galaxies in particular; a KS test rejects the null hypothesis
that the mass functions are drawn from the same underlying population
with $p_{\text{KS}}=1.4\times10^{-3}$.

For the purpose of estimating timescales (see
Section~\ref{sec:timescales}) we also evaluate the mass functions of
those PAS1 galaxies which were quenched during the epoch $0.5<z<1.0$
(based on mean stellar age from SC fits). This sub-population is shown
in magenta (clusters) and cyan (field) in the lower-left panel of
Fig.~\ref{fig:massfunction1}. We find that cluster galaxies satisfying
this condition are systematically less massive than when the whole
sample was employed. This means that the most recently quenched
objects are mostly low-mass galaxies, and the most massive PAS
galaxies were likely to have been in place already by $z=1$. This sharpens the
apparent difference between cluster and field PAS1 galaxies, based on a KS test
($p_{\text{KS}}=1.4\times10^{-13}$).

Summarising this section, we find an excess of low-mass galaxies among the PAS, PSB and SF3 populations in clusters. In contrast, we find that galaxies with high sSFR (SF1 and SF2) are suppressed in such environments. Additionally, the quenching of high sSFR galaxies in clusters seems to be mass-dependent, affecting low-mass galaxies more efficiently than massive systems.

\begin{table}
\centering
\caption{Schechter parameters of all 9 galaxy population mass functions. We use single Schechter functions except for the cluster PSBs. $M^*$ units are given in solar masses and $\phi^*$ in $\text{Mpc}^{-3}\text{dex}^{-1}$. The variable $\xi$ represents the relative change in normalisation of a cluster with respect to the field. The last two entries (*) correspond to the mass functions of galaxies quenched at $0.5<z<1.0$, while the rest correspond to the entire sample.}
\label{tab:SchechterParam}
\begin{tabular}{clccccc}
\hline\hline
     &                   & Cluster                   & Field       &       \\ \hline
     & $\alpha$ 	 &$ -1.310 \pm 0.010  $      & $ -1.402 \pm 0.006 $\\
SFT  & $\log M^*$	 &$ 10.914 \pm  0.025 $      & $ 10.930 \pm 0.010 $\\
     & $\log \phi^*$	 &$ (-3.140 \pm  0.003)\xi $ & $ -3.118 \pm 0.002 $\\ \hline
     & $\alpha$ 	 &$ -0.170 \pm  0.022 $      & $ 0.183  \pm 0.013 $\\
PAS  & $\log M^*$     	 &$ 10.787 \pm  0.015 $      & $ 10.633 \pm 0.006 $\\
     & $\log \phi^*$     &$ (-2.455 \pm  0.056)\xi $ & $ -2.699 \pm 0.032 $\\ \hline
     & $\alpha_1$  	 &$ -1.493 \pm  0.113 $      & $ -1.378 \pm 0.027 $\\
     & $\log M^*$        &$ 9.789  \pm  0.071 $      & $ 10.903 \pm 0.039 $\\
PSB  & $\log \phi^*_1$   &$ (-3.624 \pm  0.033)\xi $ & $ -4.879 \pm 0.009 $\\
     & $\alpha_2$ 	 &$ 2.448  \pm  0.297 $      &                     \\
     & $\log \phi^*_2$   &$ (-4.902 \pm  0.053)\xi $ &                     \\ \hline
     & $\alpha$  	 &$ -0.804 \pm  0.047 $      & $ -1.448 \pm 0.020 $\\
SF1  & $\log M^*$        &$ 9.334  \pm  0.020 $      & $ 9.539  \pm 0.010 $\\
     & $\log \phi^*$     &$ (-2.653 \pm  0.002)\xi $ & $ -2.444 \pm 0.006 $\\ \hline
     & $\alpha$  	 &$ -0.739 \pm  0.029 $      & $ -0.726 \pm 0.015 $\\
SF2  & $\log M^*$        &$ 10.108 \pm  0.022 $      & $ 10.125 \pm 0.009 $\\
     & $\log \phi^*$     &$ (-2.892 \pm  0.017)\xi $ & $ -2.745 \pm 0.009 $\\ \hline
     & $\alpha$   	 &$ -0.192 \pm  0.028 $      & $ 0.103  \pm 0.016 $\\
SF3  & $\log M^*$      	 &$ 10.546 \pm  0.017 $      & $ 10.462 \pm 0.007 $\\
     & $\log \phi^*$     &$ (-2.688 \pm  0.063)\xi $ & $ -2.745 \pm 0.067 $\\ \hline
     & $\alpha$    	 &$ -0.859 \pm  0.025 $      & $ -0.286 \pm 0.026 $\\
PAS1 & $\log M^*$      	 &$ 10.659 \pm  0.024 $      & $ 10.473 \pm 0.014 $\\
     & $\log \phi^*$     &$ (-3.291 \pm  0.013)\xi $ & $ -3.394 \pm 0.039 $\\ \hline
     & $\alpha$  	 &$ 0.393  \pm  0.036 $      & $ 0.488  \pm 0.025 $\\
PAS2 & $\log M^*$	 &$ 10.488 \pm  0.018 $      & $ 10.466 \pm 0.008 $\\
     & $\log \phi^*$     &$ (-2.864 \pm  0.040)\xi $ & $ -3.130 \pm 0.022 $\\ \hline
     & $\alpha$ 	 &$ 0.640  \pm  0.038 $      & $ 1.082  \pm 0.027 $\\
PAS3 & $\log M^*$      	 &$ 10.704 \pm  0.016 $      & $ 10.564 \pm 0.007 $\\
     & $\log \phi^*$     &$ (-2.746 \pm  0.026)\xi $ & $ -3.197 \pm 0.011 $\\ \hline \hline
     & $\alpha_1$  	 &$ -1.616 \pm  0.282 $      & $ -2.010 \pm 0.035 $\\
     & $\log M^*$     	 &$ 9.547  \pm  0.113 $      & $ 10.984 \pm 0.081 $\\
PSB* & $\log \phi^*_1$   &$ (-3.544 \pm  0.076)\xi $ & $ -6.145 \pm 0.008 $\\
     & $\alpha_2$ 	 &$ 1.549  \pm  0.437 $      &                     \\
     & $\log \phi^*_2$   &$ (-4.902 \pm  0.123)\xi $ &                     \\ \hline
     & $\alpha$ 	 &$ -1.253 \pm  0.022 $      & $ -0.071 \pm 0.039 $\\
PAS1*& $\log M^*$     	 &$ 10.792 \pm  0.027 $      & $ 10.477 \pm 0.017 $\\
     & $\log \phi^*$     &$ (-3.765 \pm  0.008)\xi $ & $ -3.760 \pm 0.239 $\\ \hline\hline
\end{tabular}                               
\end{table}

\subsection{Radial distribution of galaxies in clusters}
\label{sec:radialplots}                         
                                            
The radial distribution of different galaxy populations in clusters can, in principle, provide 
information on where quenching is taking place and the likely timescales.                                  
We define the centre of a cluster as its centre of mass and measure
projected distances to all galaxy members within 1~Mpc.  Additionally,
clusters are split in two richness bins ($20<N_{\text{FoF}}<45$ and
$N_{\text{FoF}}>45$ members) to reduce the influence due to variation in size, 
and stacked together to produce radial profiles.
                                            
\begin{figure*}                             
	\begin{center}                             
		\includegraphics[width=0.9\textwidth]{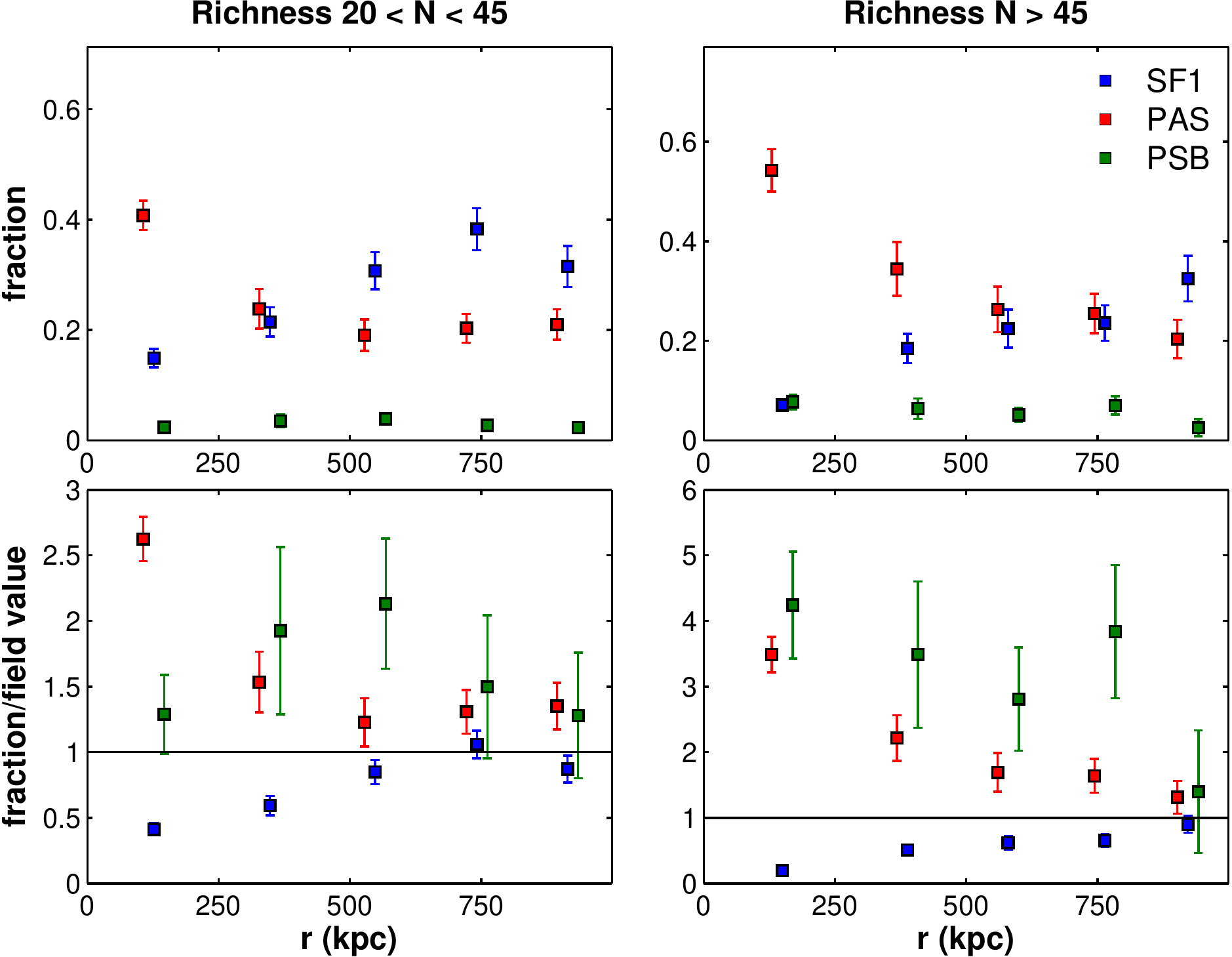}
		\vspace{-5pt}                            
		\caption{Radial plots of SF1, PAS and PSB galaxies in two cluster richness bins: clusters with between 20 and 45 and with more than 45 FoF selected members. In the top row the fraction of each population is represented as a function of cluster-centric distance. In the bottom row the fraction is normalised by the corresponding value in the field. }
		\label{fig:radialplot}                           
	\end{center}                               
\end{figure*}                               
                                            
\begin{figure*}                             
	\begin{center}                             
		\includegraphics[width=0.9\textwidth]{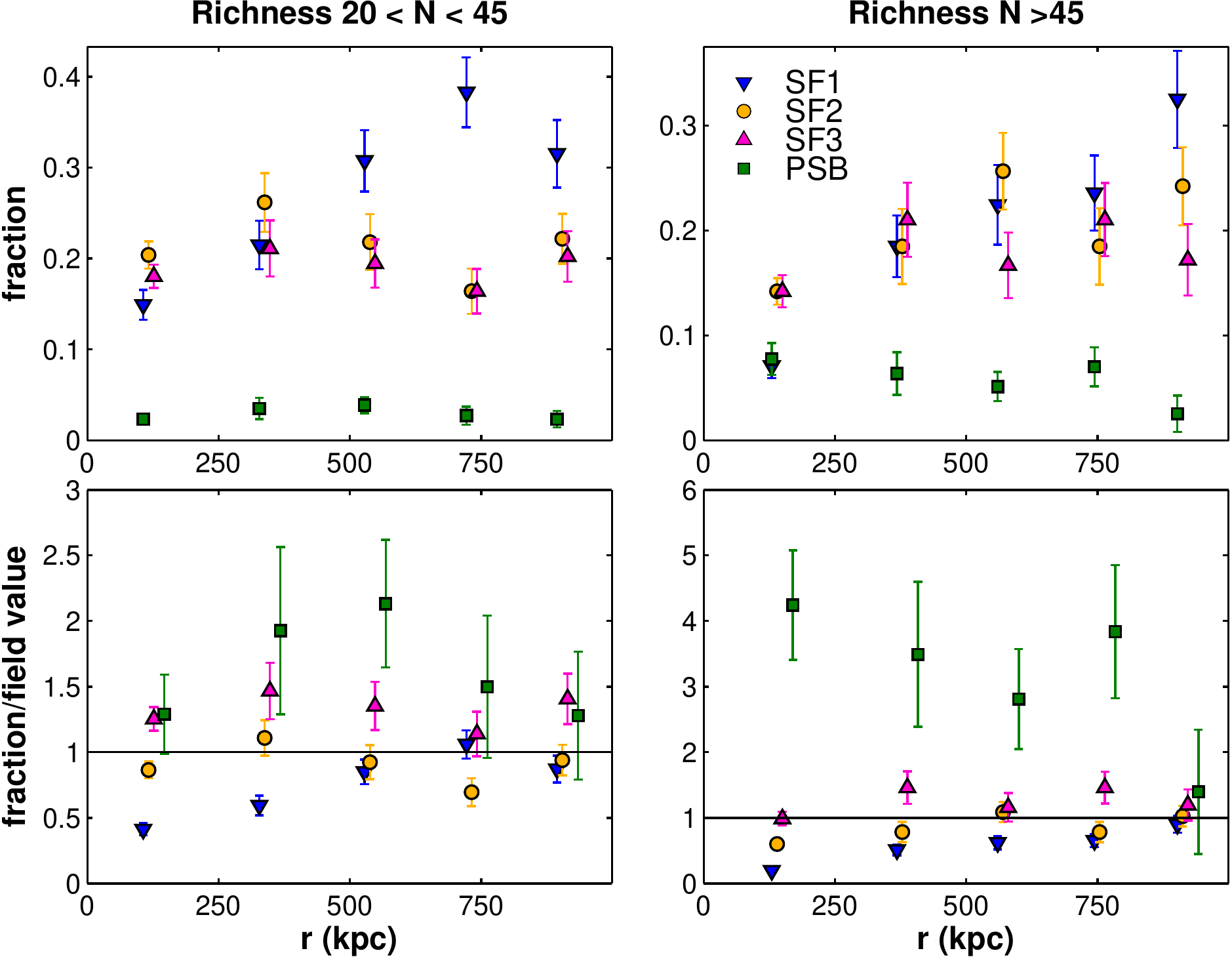}
		\vspace{-5pt}                            
		\caption{Radial plots of SF1, SF2 and SF3 galaxies in two cluster richness bins: clusters with more than 20 and fewer than 45 and clusters with more than 45 FoF selected members. In the first row the fraction of each population is represented while in the second one the fraction is normalised by the value in the field.}
		\label{fig:radialplotSF}                          
	\end{center}                               
\end{figure*}                               
                                            
The radial trends of all PAS galaxies, PSBs and SF1s are shown in Fig.~\ref{fig:radialplot}. We plot only SF1 instead of the total SF population because, as the mass functions demonstrated, this population has the strongest environmental dependence.
                                            
The radial plots show the expected trends for the star-forming and
quiescent galaxies. As in previous studies, red passive galaxies tend
to reside in the inner, denser regions of the clusters while blue
star-forming galaxies prefer the outskirts and dominate at large
cluster-centric distances \citep{oemler1974, muzzin2014}. This
difference is reflected in a KS test, which gives rise to $p_{\text{KS}} =
1.2\times10^{-12}$ and $1.0\times10^{-11}$ for the low and high
richness bins, respectively. Additionally, we find that the crossover
point between the SF1 and PAS populations scales with richness, as
expected if galaxy clusters are roughly self-similar.

PSBs are found to favour the dense cluster environment, and within 500~kpc
the fraction of these galaxies is several times higher than the field.
Although PSBs do not follow a clear radial trend, a KS test applied on the radial distributions reveals that 
formally their cluster-centric distances cannot be distinguished from those of the passive population 
(Table~\ref{tab:KSradial}). There is some evidence, however, that PSBs  are not as concentrated in
the core region as PAS galaxies. This is broadly consistent with \citet{muzzin2014},
who found that PSBs reside in the inner volumes  of clusters but avoid
the very central region. We note, however, that they also showed that this trend
weakens and the PSBs mimic the distribution of quiescent galaxies when line-of-sight velocity is omitted.
                                            
\begin{table}                               
	\centering
	\caption{The p-value of a KS-test when applied to radial distributions of different populations.}
	\label{tab:KSradial}
	\begin{tabular}{lccccc}
		\hline
		    &  \multicolumn{2}{c}{\,\,\,$\quad 20<N<45$}& &  \multicolumn{2}{c}{\,\,$\quad N>45$}   \\
		    & SF1 & PSB &  & SF1 & PSB \\
		\hline
		PAS  & $4.1\times10^{-11}$ & 0.23 &   & $8.0\times10^{-11}$ & 0.69 \\
		PSB  & 0.036 &  -  &   & $1.9\times10^{-3}$ &  -  \\
		\hline
	\end{tabular}
	\vspace{-5pt}
\end{table}

The radial distributions of SF1, SF2 and SF3, shown in
Fig.~\ref{fig:radialplotSF}, show a strong dependence of sSFR with
cluster-centric radius. The population with the highest sSFR, SF1,
presents a strong radial gradient, avoiding the inner regions of
clusters. SF2s exhibit a rather flat radial profile which drops in the
innermost bins. Finally, SF3s are the only SF population whose
fraction is higher in clusters than in the field, although the profile
is flat, similar to the SF2s.

In conclusion, the radial profiles show a pattern suggesting the more
passive populations (PAS, PSB and SF3) are more common in dense
environments than in the field and prefer to inhabit small and
intermediate cluster-centric radii. In contrast, high-sSFR 
galaxies avoid the central regions of clusters.

\section{Discussion}
\label{sec:discussion}

In this work we present the following observational evidence,
indicating that dense environments have a substantial impact on galaxy
evolution in the redshift range  $0.5<z<1.0$:

\begin{enumerate}
\item There is a high abundance of low-mass passive galaxies and PSBs
  in clusters (Fig.~\ref{fig:massfunction1}),
  and a corresponding suppression of galaxies with high sSFR
  (particularly the SF1 class) compared to the field
  (Fig.~\ref{fig:massfunction1}~\&~\ref{fig:radialplot}). This general
  trend can also be seen in the distribution of galaxies in supercolour space
  (SC1 vs SC2; see Fig.~\ref{fig:SCdistribution}), which shows that
  the cluster galaxy sample is skewed towards populations with lower sSFR.
\item There are strong radial gradients of passive and star forming
  fractions with cluster-centric distance. Passive galaxies dominate
  the central region of clusters where the galaxy density is higher,
  while star-forming galaxies prefer the outskirts
  (Fig.~\ref{fig:radialplot}~\&~\ref{fig:radialplotSF}).  In
  particular, galaxies with high sSFR (SF1) show the steepest radial
  gradients.

\end{enumerate}

In the analysis below  we use the stellar mass functions to estimate the
evolutionary connection between the various galaxy populations, and in
particular the contribution due to quenching in dense environments. We
then  identify the most likely
quenching pathways, which we describe with a simple evolutionary model.

\subsection{Contributions and timescales}
\label{sec:timescales}

In this section we estimate the contribution of each
population to the descendant class due to environmental processes. 
To achieve this we match the shapes of the stellar mass functions. 
This can be done because the SC classified galaxies correspond to 
92.7\% of the total sample (the rest correspond to rarer dusty, 
metal-poor or non-classified galaxies), so we  assume that they evolve
from one population to another without missing a significant fraction.

In the absence of enhanced quenching processes, we may consider a
``slow fading'' route, driven by the gradual decay of sSFR as galaxies
build up stellar mass, which qualitatively agrees with the observed
shift towards higher mass as galaxies age (see
Fig.~\ref{fig:massfunction1}).  In contrast, environmental processes
are thought to act rapidly \citep{muzzin2012, wetzel2012, wetzel2013,
  mok2013}, so that galaxies do not build up a significant amount of
stellar mass in the process of being quenched. In this scenario,
galaxies migrate to a different population while the shape of the
original mass function remains unchanged. Therefore, there are two
processes that contribute to the build up of the cluster mass function
according to this simple evolutionary scheme; accretion of field galaxies of the
same type, and injection of galaxies from other populations due to the
action of the environment. Consequently, some cluster mass functions
are composites of other populations, while this is not the case in the
field.

We estimate the composition of the cluster populations by fitting each
stellar mass function with a simple model (see Equation
\ref{eq:model}), consisting of a linear combination of other
populations \footnote{As an important caveat, we note that this model 
does not allow for effects of merging, which would imply evolution 
from one population to another with a significant change in stellar mass.},
\begin{equation}
\label{eq:model}
 \tilde{\phi}_{\text{Cluster}}^{i} = \alpha \phi_{\text{Field}}^{i} + \sum_j \beta_j \phi_{\text{Field,Cluster}}^{j}
\end{equation}
where $\phi$ are the various galaxy mass functions. The subindex $i$
corresponds to the population we are trying to model and the subindex
$j$ to all the possible contributors. The terms $\alpha$ and $\beta$
represent the relative contributions of the progenitor
classes  to the target population.  The fitting
is conducted  using a Monte-Carlo method, minimising $\chi^2$ 
while the data points are allowed to vary within errorbars.

The key assumption made when using equation~\ref{eq:model} is that
quenched galaxies do not experience rejuvenation, i.e. there is no
flow of galaxies from PAS and PSB populations towards the SF class, or from
PAS galaxies into PSBs.  Additionally, PSBs and SF3s are the only
populations that share a boundary with the passive sequence (in the
SC-diagram, see Fig.~\ref{fig:SCdiagram}). Hence, in order to become
passive a galaxy must evolve across this boundary.  Therefore we only
consider these two populations as contributors to the PAS
populations. We assume the field SF mass function is the population
being quenched, i.e. we assume these galaxies are quenched when they
enter a cluster environment.

\begin{table}
\centering
\caption{The estimated contribution to the cluster galaxy populations
  (1st column) from the progenitor classes, based on fitting the
  galaxy mass functions (see Equation~\ref{eq:model}).  Contributions
  are expresed as fractions of the progenitor and the target
  populations. Those entries marked with (f) correspond to the field,
  otherwise they represent cluster populations. The third column
  corresponds to the contribution relative to the progenitor
  populations, while the fourth column represents the fraction of the
  final population that  comes from each progenitor class.}
\label{tab:parameters}
\begin{tabular}{llll}
\hline
                       & $\Phi$   & Contribution              & \%              \\ \hline
\multirow{4}{*}{PSB*}  & SF1(f)   & $\beta  = 0.11\pm0.01   $ & $96.1\pm7.1\%     $ \\
                       & SF2(f)   & $\beta \sim 10^{-4}     $ & $<1\%         $ \\
                       & SF3(f)   & $\beta \sim 10^{-5}     $ & $<0.1\%       $ \\
		       & PSB*(f)  & $\alpha = 0.23\pm0.04   $ & $3.8\pm0.7\%  $ \\ \hline
\multirow{3}{*}{SF3}   & SF1(f)   & $\beta  = 0.013\pm0.005 $ & $<1\%         $ \\
                       & SF2(f)   & $\beta  = 0.12\pm0.04   $ & $12.6\pm3.7   $ \\
                       & SF3(f)   & $\alpha = 1.4\pm0.2     $ & $87.1\pm3.8\% $ \\ \hline
\multirow{3}{*}{PAS1$^*$} & SF3      & $\beta  = 0.22\pm0.02   $ & $26.6\pm3.1\% $ \\
                       & PSB      & $\beta  = 3.41\pm0.45   $ & $73.3\pm3.0   $ \\
                       & PAS1*(f) & $\alpha = 0.02\pm0.01   $ & $<1\%         $ \\ \hline
\end{tabular}
$^*$~Galaxies quenched at $0.5<z<1.0$ selected using mean stellar age information.
\end{table}

No assumption is made regarding the progenitors of cluster PSBs, hence
all SF and field PSBs are considered potential candidates and
introduced in Eq.~\ref{eq:model}.  We find that the shape of the
cluster PSB mass function is recovered if $96.1\pm7.1\%$ 
of its galaxies are accreted from the
SF1 class and $3.8\pm0.7\%$ are accreted from the field PSB
population. This is because field SF1 galaxies are the only population with a
similar shape to cluster PSBs, i.e. steep at the low-mass end. The
contributions from SF2s and SF3s are less than 1\% (see
Table~\ref{tab:parameters}).

We also include all the SF populations in order to reproduce the
cluster SF3 mass function. We find that the excess of low-mass SF3s in
clusters is reproduced by adding a contribution solely from the SF2
population, with $12.6\pm3.7\%$ of cluster SF3s evolving from field
SF2s, while accretion from field SF3s accounts for the remaining
$87.1\pm3.8\%$. The field SF1 mass function does not provide a good
fit to the cluster SF3 mass function, implying that essentially all
environmentally quenched SF1s evolve through the PSB route.

In order to estimate the visbility time of the PSB phase, we first
apply the analysis to the subset of the younger PAS1 galaxies that were
quenched over the redshift range $0.5<z<1.0$ (magenta and cyan lines
in Fig. \ref{fig:massfunction1}).  These galaxies are selected at a
given redshift based on their mean stellar age, as obtained from the
SC fitting procedure (see Section~\ref{sec:classification}). As
mentioned previously, we only consider cluster PSBs, cluster SF3s and
field PAS1 as potential progenitors for the PAS galaxies.  The
similarity in shape of the cluster SF3 and field PAS1 mass functions
does lead to some degenaracy affecting the contributions of these
populations. This does not affect the contribution from
PSBs, however. We find that $73.3\pm3.0\%$ of the cluster PAS1 population
that were quenched in the redshift range $0.5<z<1.0$ come from cluster PSBs
(with the remaining $26.6\pm3.1\%$ from cluster SF3s).

We use these contributions to estimate the visibility timescale
($\tau_\text{vis}$) for the PSB phase. The redshift range $0.5<z<1.0$
corresponds to a time interval $\Delta t = 2.7\pm0.3$~Gyr.  The
visibility timescale is calculated dividing $\Delta t$ by the expected
number of times the observed PSB population has evolved into  PAS1 galaxies
during this time interval (i.e. $\beta_{PSB}$).
\begin{equation}
\label{eq:timescale}
\tau_{\text{vis},j} = \frac{\Delta t}{\beta_j}
\end{equation}

Expressed in terms of the parent population, the PSB
contribution to PAS1s corresponds to $3.41\pm0.45$ times the observed number of
PSBs in clusters. This means that more than three times the current
number of these galaxies must have faded into the red sequence over a
time period of $\sim2.7$~Gyrs. Therefore, the visibility time for PSBs
is $0.8\pm0.1$~Gyrs.

In Section~\ref{sec:PSBinclusters} we explore the visibility time
for the PSB phase from a theoretical perspective, using stellar
population synthesis models \citep{wild2016}. These simulations
estimate visibility times between $0.4$--$1$~Gyrs, consistent with the
estimates obtained using stellar mass functions.

\subsection{Evolutionary pathways}
\label{sec:evol.paths}

We now develop a simple evolutionary model to link the various populations considered in this paper.

\begin{figure}
	\begin{center}
		\includegraphics[width=0.45\textwidth]{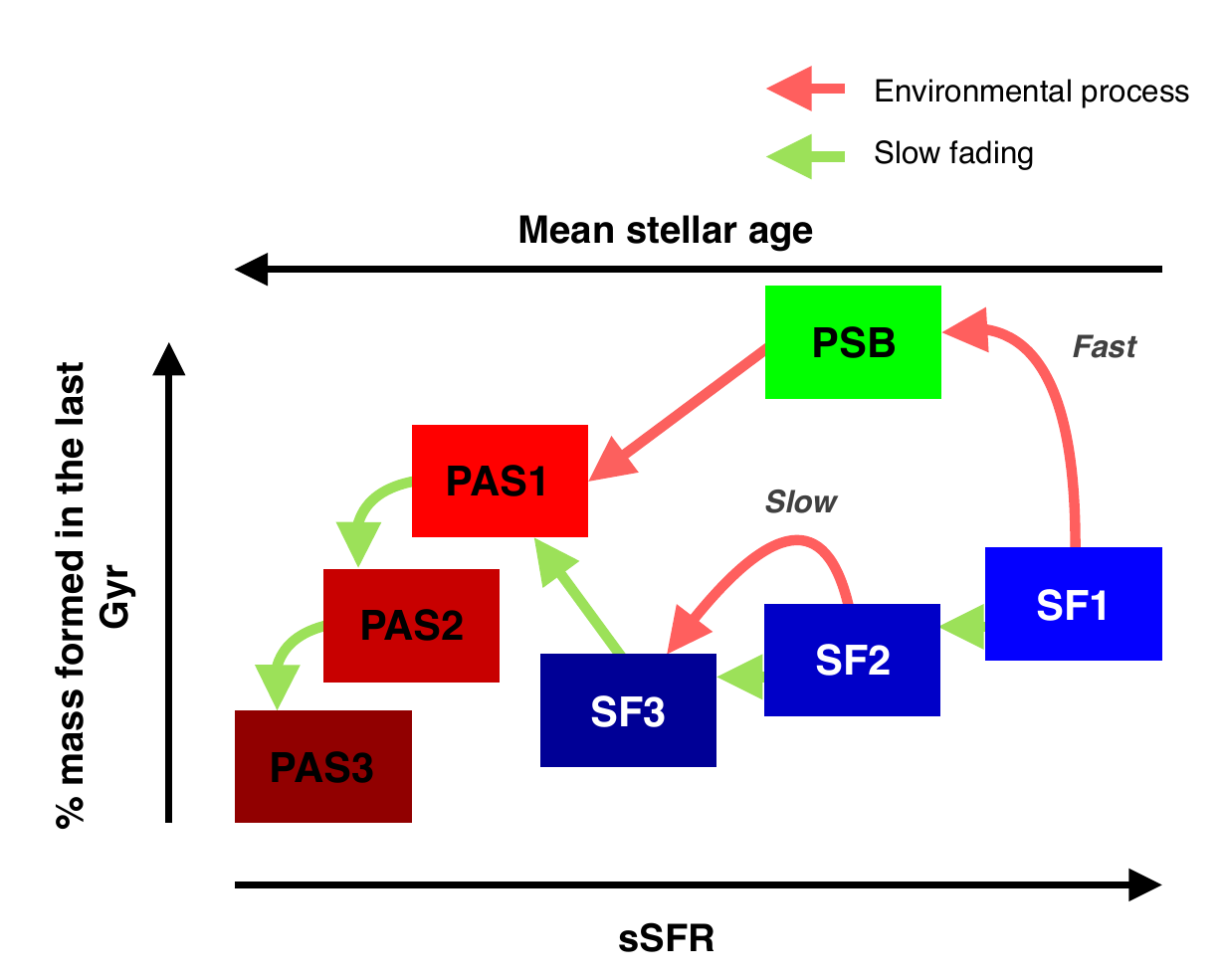}
		\vspace{-10pt}
		\caption{Scheme of our proposed evolutionary pathways. Green arrows illustrate the evolution of galaxies with constant SFR while the red arrows represent evolution driven by environment i.e.\ SFR being truncated by some environmental mechanism.}
		\label{fig:evolcartoon}
	\end{center}
\end{figure}

We assume that the  evolution of low-mass galaxies ($M<10^{10.5}~\text{M}_\odot$)
in the field at $z<1$
is mainly dominated by slow, undisturbed evolution. An isolated
star-forming galaxy builds up stellar mass so that the sSFR drops and
the galaxy slowly fades and moves through the star-forming classes (SF1, SF2, SF3)
to eventually become passive (PAS). This slow fading is shown by the green arrows in
Fig.~\ref{fig:evolcartoon}. In order to produce the bulk of the PSB
population additional (rapid) quenching mechanisms are needed.

We suggest that the cluster environment  causes the
deviations from the slow fading path. Based on the contributions 
calculated in Section~\ref{sec:timescales}, we conclude that this can happen in two ways.
Rapid quenching affects galaxies  with very high sSFR (SF1), which  are quenched rapidly during infall,
giving rise to PSBs. This explains the sharp upturn of the PSB
stellar mass function at the low-mass end, which matches the field
SF1  mass function. 
Secondly, galaxies with
intermediate sSFRs (SF2) may also  be quenched, causing them to prematurely evolve into SF3
galaxies.  These environmentally-driven paths are represented with red
arrows in Fig.~\ref{fig:evolcartoon}. After quenching has taken place
all galaxies converge to the quiescent population, regardless of the
quenching pathway they followed. First they evolve to the youngest
passive population (PAS1), then progressively evolve into PAS2 and
PAS3 as they age and/or dry-merge.

We now analyse the insight provided by the radial distributions, which in
principle can probe the location of the environmental quenching and
constrain the likely timescales. The SF1 population is found to be
strongly depleted in the cluster core; a KS test confirms its distribution is 
inconsistent with a flat distribution ($p_\text{KS}\sim10^{-5}$). This implies that the timescale for
this quenching process is short, and less than the typical dynamical timescale of
clusters ($<10^9$~years), as otherwise the radial trend would dilute.
In contrast, neither the SF2 or SF3 populations show strong radial
trends ($p_\text{KS}\sim0.24$). Therefore the second
evolutionary path must be a more gradual process and take longer than
the dynamical timescale, i.e. $\gtrsim 10^9$~years.

Finally, we note that PSBs show no strong radial gradients, which
implies that either environmental quenching occurs everywhere within
the inner Mpc of the cluster, or the visibility time of the PSB phase
is comparable to the dynamical timescale, $\sim~1$~Gyr. As noted
above, however, the quenching timescale to convert SF1 galaxies into
PSBs must be considerably shorter.

In summary,  our results suggest more than one quenching mechanism acting in clusters,
which seem to act on different timescales.  One of them preferentially
influences low-mass galaxies with high sSFR, while a second quenches
galaxies with intermediate sSFRs.

\subsection{PSB in clusters and the field}
\label{sec:PSBinclusters}

The properties of PSB galaxies within clusters differ from PSBs in the
field: their distribution in SC1--SC2 space is different as well as
their mass functions. This suggests PSB galaxies may be produced
through different processes depending on the environment.

\begin{figure}
	\begin{center}
		\includegraphics[width=0.45\textwidth]{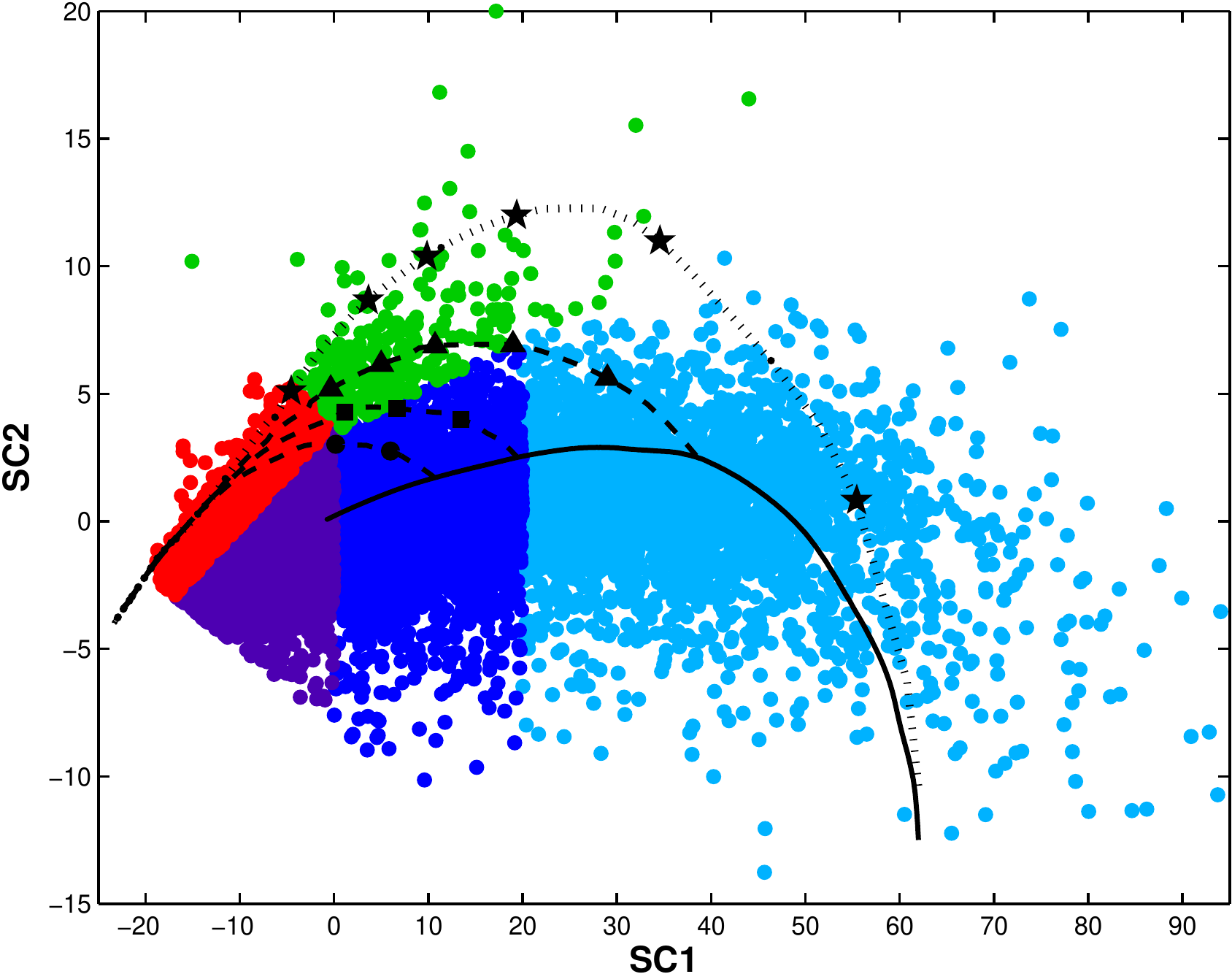}
		\vspace{-5pt}
		\caption{Evolutionary tracks in supercolour space, based on the \citet{bruzual2003} models used in \citet{wild2016}. Filled circles represent the galaxies in our sample and their colours correspond to the population they belong to (Fig.~\ref{fig:SCdiagram}). The solid line traces the evolution with constant SFR. The dotted line represents an exponentially decaying SFR with a timescale of 0.1~Gyr. Dashed lines correspond to continuous SFR and exponential truncation (with a timescale of 400~Myr) of the star formation at different times: 1, 3 and 6~Gyr after formation. Black symbols mark intervals of 0.2~Gyr starting when the SFR first drops.}
		\label{fig:SCtracks}
	\end{center}
\end{figure}

To analyse the possible origins of PSB galaxies we use the stellar
population synthesis models presented in \citet{wild2016}. These
models consist of three different star formation histories (SFH; see
Fig.~\ref{fig:SCtracks}): (1) with constant SFR, corresponds to
unaltered evolution and a gradual drop in sSFR (solid line); (2)
exponentially declining SFH with a decay time of 100~Myrs,
representing galaxies that have undergone a strong burst of star
formation that is rapidly truncated due to depletion of the gas
reservoir (dotted line); and (3) exponential truncation of star
formation with decay time of 400~Myrs after an extended period of
continuous star formation of 1, 3 and 6~Gyrs since formation (dashed
lines). In our case, this rapid truncation is assumed to be the effect
of environmental quenching.

We see that the last two SFHs can lead to a PSB phase.  In either case
the maximum value of SC2 reached depends on  the rapidity of
the quenching event and the fraction of the stellar mass built up
during the last Gyr. Hence PSBs formed immediately after a starburst
event systematically reach higher values of SC2 than PSBs which were
quenched after a more extended episode of star formation.

The distribution of PSBs in the SC diagram (Fig.~\ref{fig:SCdistribution} described in 
Section~\ref{sec:quenchinginclusters}) suggests that PSBs are triggered by different mechanisms in different environments. In particular, those PSBs in clusters
are unlikely to be produced after a significant starburst, in which the galaxy formed a considerable fraction of its stellar mass. Instead, they are more likely to have originated via rapid quenching after an extended period of star formation or after a more marginal burst of star formation. We find that PSBs in clusters are concentrated at $\text{SC2}<10$ while in the field they reach much higher values
($\text{SC2}\sim15$; see Fig.~\ref{fig:SCdistribution}). In addition,
this quenching must be fast ($\tau_{\text{Q}}\sim400$~Myrs, from
simulations) to cause a galaxy to leap off the slow evolution path into the
PSB regime. With much longer SFR decay times the evolution would be
indistinguishable from the undisturbed case.  This matches the
quenching timescale $<1$~Gyr suggested by the radial gradient of SF1
galaxies in clusters.

Additionally, the models show that the visibility time of the PSB
phase is longer if a higher value of SC2 is reached.  Hence, those
preceded by a starburst tend to have longer visibility times than
those produced by rapid truncation after more extended star formation. Similarly, if the episode of star
formation carries on for too long before being  truncated, the galaxy
will not reach the PSB regime at all. These two factors constrain the
value of the PSB visibility timescale to the range $0.4<\tau_\text{vis}<1$~Gyr.

In conclusion, PSBs in galaxy clusters are more likely to be produced
via rapid truncation after an extended period of star
formation or after a minor starburst rather than being the result of a major
starburst. Simulations confirm, regardless of the underlying process,
that the quenching  must act quickly to  produce the PSB
imprint, otherwise galaxies would stay too close to the undisturbed
evolutionary pathway.

\subsection{Mechanisms that can cause fast- and slow-quenching}
\label{sec:potentialmechanisms}
Our results suggest that cluster galaxies at $0.5<z<1$ quench via at least two
different pathways. A single mechanism may be responsible, which
affects galaxies differently depending on their properties,  or
several quenching mechanisms may act simultaneously to produce the
different evolutionary sequences.

One pathway, which we refer to  as \lq fast-quenching\rq, acts on short
timescales, quenching galaxies faster than a cluster dynamical
time. It predominantly affects galaxies with high sSFRs and is more
efficient at quenching low-mass galaxies.  It becomes significant at
cluster-centric radii $R\lesssim750$~kpc. The other pathway, which we
label \lq slow-quenching\rq, acts on longer timescales, comparable to
or greater than the cluster dynamical timescale
($\tau_{\text{slow}}~\gtrsim~1$~Gyr). Slow quenching predominantly affects galaxies
which exhibit moderate sSFRs, and shows no trend with stellar mass nor
cluster-centric radius.

We consider it unlikely that the enhanced quenching in clusters is
produced by internal galaxies processes, such as AGN or stellar
feedback. Powerful AGN feedback is generally believed to occur in massive galaxies,
so it is unlikely to cause the fast-quenching described above, which
is more efficient at quenching low mass galaxies. Furthermore,
star-formation-driven winds are also unlikely to be the  primary cause, 
as Fig.~\ref{fig:SCtracks} shows no evidence for strong starbursts in
cluster galaxies.

The main contenders for the mechanisms responsible for fast- and
slow-quenching are interactions between the ICM and galaxies (such as
ram pressure stripping and strangulation), and galaxy-galaxy
interactions (such as harassment, mergers and tidal interactions).


Ram pressure stripping of the cold gas reservoir within a galaxy can
quench star formation in a few hundred Myrs
\citep{steinhauser2016}. This mechanism acts preferentially in the
central region of galaxy clusters or groups \citep{rasmussen2006, kawata2008}, where the ICM is densest
and galaxies have high velocities. Furthermore, ram pressure stripping
removes the cold gas reservoirs of low-mass galaxies more efficiently
than high-mass galaxies as their lower gravitational potential is
unable to keep the gas bound against the ram pressure. These characteristics
can produce the observed properties of the fast-quenching mode
described above, so ram pressure stripping is one of the contenders
for causing the fast-quenching in clusters.

Galaxy mergers may also quench galaxies quickly. A merger can funnel
gas into the centre of a galaxy, triggering a nuclear burst of star
formation that may deplete the gas reservoir in a fraction of a
Gyr. Although the merger cross section is small in the centre of
clusters \citep{ostriker1980, makino1997}, these encounters frequently
occur in cluster outskirts, as well as in groups. Our cluster sample
is likely to have a broad range of velocity dispersions. By comparing
our sample with the X-ray sample from \citet{finoguenov2010} we
estimate the majority of our structures have velocity dispersions of
$\sigma_{\text{v}}=300-500~\text{kms}^{-1}$, so mergers may be
frequent. 
However, the only type of merger able to produce the PSB stellar mass distribution is a major merger between two low-mass galaxies (i.e. two SF1s) and the resulting starburst would cause a high value of SC2, that is inconsistent with the typical values of SC2 found in cluster PSBs. Therefore, some external mechanism (e.g. gas stripping by ICM) may be required to decrease the gas fraction present in these galaxies in order to prevent a major starburst from occuring.

Galaxy encounters which cause tidal interactions, such as galaxy
harassment, are much more frequent in groups and clusters than
mergers, and these processes can strip gas from galaxies and reduce
their  SFR. Due to the high relative velocities of galaxies in clusters,
these interactions are too quick and inefficient to be the direct
cause of fast-quenching evolution \citep{boselli2006, byrd1990}, but
they may be responsible for slow-mode quenching.

At this point we are unable to pinpoint the mechanism that produces
the fast-quenching within $0.5<z<1$ clusters. However, future studies
of the morphology of cluster PSBs may shed some light on which
mechanism is responsible. Mergers would produce PSBs with disturbed/spheroidal 
morphologies, as the interaction disrupts the structures of the
galaxies, whilst ram pressure stripping/strangulation would result in PSBs with
more disc-like morphologies, as the galaxy would quench before the
disc fades.

Many of the features exhibited by the slow-quenching mechanism can be
explained by galaxy strangulation, where the hot gas envelope of the
galaxy is removed by the ICM. For example, strangulation halts star
formation gradually over $\sim4$~Gyrs \citep{bekki2002}. The hot gas
reservoir of a galaxy is easily removed through interactions with the
ICM, therefore strangulation affects both high and low-mass galaxies
equally.

However, there are other potential processes responsible for
slow-quenching. Galaxy harassment, as mentioned before, could
significantly affect the star formation of a galaxy after a number of
encounters, which requires a few Gyrs. Similarly, mergers involving
galaxies with low gas content and intermediate sSFRs (SF2) may quench
galaxies without following the PSB route.

\section{Conclusions}
\label{sec:conclusions}

We have optimised a Friends-of-Friends algorithm to find galaxy
overdensities in the UKIDSS UDS field, allowing us to analyse the
relationship between environment and galaxy quenching. In the redshift
range $0.5<z<1.0$ we identify 37 candidate galaxy clusters containing
at least 20 galaxies.  To analyse the field and cluster galaxy
populations, we use the PCA galaxy classification scheme of
\citet{wild2016}, which allows us to separate star-forming, passive,
and recently-quenched ``post-starburst'' (PSB) galaxies using
photometric data.  Comparing the resulting stellar mass functions, and
the radial distributions for cluster populations, our key findings can be summarised as follows:

\begin{enumerate}

\item We find
  evidence for an overabundance of low-mass passive galaxies and PSBs
  in galaxy clusters compared to less dense environments. The PSB
  population show a very steep stellar mass function in clusters, dominated by galaxies at
  low mass ($M<10^{10}~\text{M}_\odot$).

\item Galaxy clusters show a relative underabundance of galaxies with
  high specific star-formation rates (SF1 galaxies). 
The SF1 mass
  function is steep, suggesting that rapid quenching of this
  population in dense environments provides a natural explanation for the corresponding
  excess of PSBs.

\item 
The radial distribution of galaxy types reveals a decline in the
fraction of star-forming galaxies towards cluster cores, with a
corresponding steep rise in the passive galaxy population. The SF1
population  show a very steep decline towards cluster cores, suggesting
very rapid quenching of these galaxies on entering dense environments,
on a timescale less than the cluster dynamical timescale ($<1$~Gyr).

\item We measure a typical visibility time for the PSB phase of
  galaxies within clusters of $800\pm100$~Myrs, based on a
  comparison of stellar mass functions.

\item We find that PSBs in galaxy clusters are most likely to be produced by
  a rapid truncation following an extended period of star formation or after a minor starburst,
  rather than gas depletion after a major starburst. This may imply that environmental mechanisms typically quench
  galaxies without triggering any significant burst of star formation.

\end{enumerate}

To explain the relative abundances and radial distributions, we
suggest there are two main quenching pathways occurring in clusters:
rapid quenching and slow quenching.  The first path affects galaxies
with high sSFR (SF1), predominantly at low mass, 
which quench rapidly to become PSBs and
thereafter build up the low-mass end of the passive red sequence.
The second pathway
affects star-forming galaxies with moderate sSFR (SF2), accelerating
their decay in sSFR over an extended period of time, comparable to the
dynamical timescale of a galaxy cluster.

The processes behind fast environmental quenching need to act on
timescales shorter than $1$~Gyr, quench preferentially high
sSFR/low-mass galaxies, and produce a strong radial dependence without
inducing a strong starburst.  Ram-pressure stripping provides a likely
explanation, although we cannot rule out a contribution from other
processes (such as merging).
  Similarly, the processes behind slow quenching act on
timescales comparable to the cluster dynamical time or longer,
affecting galaxies with intermediate sSFR regardless of their stellar
mass. Such trends can be explained through strangulation, gradual galaxy harassment, or gas-poor mergers.

In summary, we conclude that environmental processes appear to have a significant impact on the
properties of low-mass galaxies in the redshift range $0.5<z<1.0$.

\section{Acknowledgements}

This work uses data from ESO telescopes at the Paranal Observatory
(programmes 094.A-0410 and 180.A-0776; PI: Almaini). We are grateful
to the staff at UKIRT for their tireless efforts in ensuring the
success of the UDS project. We also wish to recognize and acknowledge
the very significant cultural role and reverence that the summit of
Mauna Kea has within the indigenous Hawaiian community. We were most
fortunate to have the opportunity to conduct observations from this
mountain. MS acknowledges support from IAC and STFC. VW acknowledges 
support from the European Research Council Starting grant (SEDmorph, P.I. V. Wild). 
We also use data from the VIMOS Public Extragalactic Redshift Survey (VIPERS). VIPERS
has been performed using the ESO Very Large Telescope, under the
"Large Programme" 182.A-0886. The participating institutions and
funding agencies are listed at http://vipers.inaf.it.

\bibliographystyle{mnras} 
\bibliography{biblio}


\bsp	
\label{lastpage}
\end{document}